\documentclass[prc,showpacs,showkeys]{revtex4}
\usepackage{epsfig,graphicx,float,color}
\usepackage{amssymb}
\usepackage[normalem]{ulem}

\begin{document}

\title{Nuclear shape coexistence in Po isotopes: An interacting boson model study}

\author{J.E.~Garc\'{\i}a-Ramos$^1$ and K.~Heyde$^2$}
\affiliation{
$^1$Departamento de  F\'{\i}sica Aplicada, Universidad de Huelva,
21071 Huelva, Spain\\
$^2$Department of Physics and Astronomy, Ghent University,
Proeftuinstraat, 86 B-9000 Gent, Belgium}
\begin{abstract} 
\begin{description}
\item [Background:] The lead region, Po, Pb, Hg, and Pt,  shows up the presence of
coexisting structures having different deformation and  corresponding
to different particle-hole configurations in the Shell Model language.   

\item [Purpose:] We intend to study the importance of configuration
  mixing in the understanding of the nuclear structure of even-even Po isotopes,
  where the shape coexistence phenomena are not clear enough. 
 
\item [Method:]  We study in detail a long chain of polonium isotopes,
$^{190-208}$Po, using the interacting boson model with configuration
mixing (IBM-CM). We fix the parameters of the Hamiltonians through a
least-squares fit to the known energies and absolute B(E2) transition
rates 
of states up to $3$ MeV.

\item [Results:] We obtained the IBM-CM Hamiltonians and we calculate
excitation energies, B(E2)'s, electric quadrupole moments, nuclear
radii and isotopic shifts, quadrupole shape invariants,  wave functions, and deformations.  
 
\item [Conclusions:] We obtain a good agreement with the experimental
data for all the studied observables and we conclude that shape
coexistence phenomenon is hidden in Po isotopes, very much as in the case of
the Pt isotopes.
\end{description}
\end{abstract}

\pacs{21.10.-k, 21.60.-n, 21.60.Fw}

\keywords{Po isotopes, shape coexistence, intruder states, energy fits.}

\date{\today}
\maketitle

\section{Introduction}
\label{sec-intro}

Shape coexistence has been observed by now all through the nuclear mass region, encompassing both light
nuclei ($^{16}$O region) \cite{morinaga56}, proceeding up to the region of very heavy nuclei in the Pb region \cite{duppen84,andrei00}, and
have been reviewed in a number of papers over a period spanning about 3 decades \cite{hey83,wood92,heyde11}.
Recent advances in experimental methods to explore nuclei, removed far from the region of $\beta$-stable
nuclei, have opened up possibilities to explore the appearance and behavior of shape coexistence in series
of isotopes and isotones \cite{blum13}. A  wealth of new data, on both energy systematics, but, more important,
on observables such as in-beam spectroscopy and lifetime data \cite{jul01,gade08}, Coulomb excitation using inverse kinematics \cite{gorgen10},
direct reactions on unstable nuclei \cite{hansen03}, radioactive decay modes at the limits of nuclear stability
\cite{pfutz12}, and break-up reactions \cite{bauman12} have been at
the basis of recognizing the rather universal 
appearance of shape coexisting phenomena \cite{heyde11}. Moreover, measurement of the essential ground-state
properties, such as masses \cite{blaum13}, charge radii \cite{cheal10} and nuclear moments \cite{neyens03}, as well
as the possibilities to study monopole E0 transitions in nuclei \cite{wood99,kib05}, helped completing the data basis in such
a way as to  confront theoretical modelling of nuclear structure properties in greater detail than before.

From a theoretical side, present-day methods starting from both the nuclear shell model, or approaching
the atomic nucleus using mean-field methods have resulted in developments of both new algorithms as well as 
making use of the increased computing possibilities (see
\cite{caurier99,brown14,shimi12} and references therein). The present
status has evolved in a situation where the conditions for shape coexistence to occur are becoming understood.
It looks like a balance between two opposing nuclear force components, i.e., on one side the stabilizing effect
caused by the presence of closed shells (the monopole part), aiming at stabilizing the nucleus into a spherical
shape, versus the low-multipole (mainly quadrupole) components redistributing protons and neutrons into a
deformed shape is at the origin of the appearance of shape coexistence in a given mass region.
Recent large-scale shell-model studies (LSSM), using diagonalization in a very large many-open shell basis
in various mass regions \cite{caurier05} or
making use of an advanced Monte-Carlo shell-model approach (MCSM) \cite{shimi12}, have been carried out.  
Besides, the concept to start from deformed average potentials and calculating the total energy curves as a function
of deformation has been explored, in particular, for nuclei in the
Pb region \cite{bengt87,bengt89,naza93}. However, recent studies
\cite{bender03}, using a microscopic approach to determine the optimal mean fields, even
going beyond when bringing in the nuclear dynamics, have given quantitative results - 
using both Skyrme forces \cite{Skyr59,Vaut72,bender03,erler11} and 
Gogny forces  \cite{gogny73,gogny75,gogny80,gogny88,liber99}, as well as making use of a relativistic mean-field approach 
\cite{walecka74,serot86,reinhard89,serot92,ring96,niksic11} - that are indicative of the above mechanism.
Moreover, attempts have been made and are still improved to extract a
Bohr Hamiltonian \cite{bomot75,rowe10} starting from a microscopic basis \cite{proc09,dela10,matsu10}.  
\begin{figure}[hbt]
  \centering
  \includegraphics[width=0.90\linewidth]{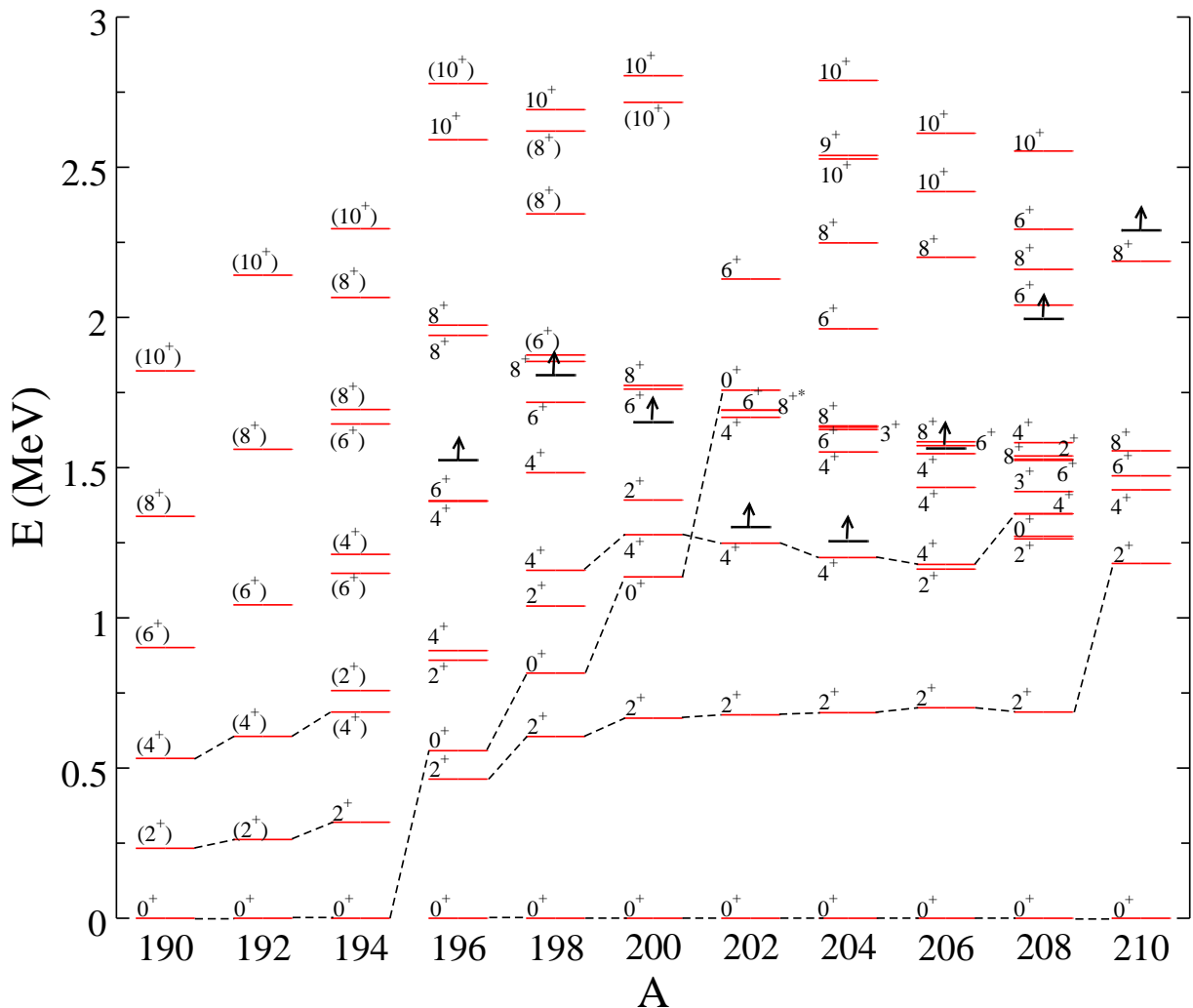}
\caption{(Color online) Experimental energy level systematics for the Po isotopes.
  Only levels up to $E_x \sim$ 3.0 MeV are shown. The symbol $\uparrow$ indicates 
the beginning of the region where extra levels of low-spin, indicated
with spin, parity assignments between brackets,
in the Nuclear Data Sheet references, start appearing. The symbol $\ast$ for the 8$^+$ level for mass A=202 indicates that the energy
is within an interval $\leq$ 40 keV above the 6$^+$ level. Dashed
lines connect states that are supposed to have a similar structure.}
  \label{fig-system-hg}
\end{figure}

From a microscopic shell-model point of view, the hope to treat on equal footing the
large open neutron shell from N=126 down to and beyond the mid-shell
N=104 region, jointly with the valence protons in the Pt, Hg, Po, and Rn
nuclei, even including proton multi-particle multi-hole (mp-nh) excitations across
the Z=82 shell closure, is beyond present computational possibilities.
The truncation of the model space, however, by concentrating on nucleon pair
modes (mainly $0^+$ and $2^+$ coupled pairs, to be treated as bosons
within the interacting boson approximation (IBM) \cite{iach87}),
has made calculations feasible, even including pair excitations
across the Z=82 shell closure \cite{duval81,duval82} in the Pb region in a transparent way.
More in particular, the Pb nuclei have been extensively studied giving rise
to bands with varying collectivity depending on the nature of the
excitations treated in the model space
\cite{hey87,hey91,fossion03,helle05,paka07,helle08}.
More recently, detailed studies of the Pt nuclei have been carried out
\cite{king98,harder97,Garc09,Garc11,Garc14a,cutcham05a,cutcham05} as  
well as for the Hg nuclei \cite{bar83,bar84,Garc14} in an attempt to describe the large amount of low-lying states and their
E2 decay properties, explicitly including particle-hole excitations across the Z=82 shell 
closure.
A novel mapping procedure to determine an algebraic IBM Hamiltonan has
been proposed by Nomura {\it et al.}~\cite{nomura08,nomura10}. In contrast
to the standard shell-model to boson model mapping method \cite{arima77, otsuka78}, it has been shown 
that an IBM Hamiltonian can be determined, mapping a self-consistent mean-field total energy surface E($\beta,\gamma$)
(over the full $\beta-\gamma$ plane) onto the corresponding IBM mean-field energy. Very recently, 
this method was extended to include intruder mp-nh configurations, with a detailed
coverage of the Pt, Pb and Hg isotopes \cite{nomura11a,nomura11b,nomura12a,nomura13}.

Recently, a lot of new experimental results have become available in the Pb region, encompassing, besides the Pb isotopes,
the nearby Hg, Pt and Po isotopes (see Section \ref{sec-exp} for references on the Po nuclei). In both the Pb and Hg isotopes,
there is overwhelming information (excitation energies, B(E2) values, isotopic shifts, $\alpha$-hindrance
factors, etc.) by now that highlights the presence of shape
coexistence, which is associated
with the presence of prolate, oblate and spherical bands in the
case of the Pb nuclei, and with the presence
of a prolate deformed band together with a less deformed oblate structure,
forming the yrast band in the case of Hg \cite{jul01}. 
Whereas the intruder bands are easily singled out for the Pb and Hg nuclei in which 
the excitation energies display the characteristic parabolic pattern with
minimal excitation energy around the N=104 neutron mid-shell nucleus, this
structure is not immediate in both the Pt and the Po nuclei.
Therefore, in the present paper, we carry out an extensive study of the Po nuclei within the context of the interacting
boson model, including 2p-2h excitations across the Z=82 proton closed
shell, thereby extending the regular IBM model space, containing N
bosons, with the intruder N+2 IBM model 
space, also taking into account the interaction between both subspaces,
which is called the IBM-CM approach.

The paper is organized as follows: in Section \ref{sec-exp}, we 
describe the experimental situation in the Po isotopes whereas in Section \ref{sec-theo}, we present the various theoretical approaches 
that have been used in the literature to study the Po nuclei; in Section \ref{sec-ibm-cm}, we succinctly present the IBM-CM
formalism as well as the fitting methodology used; here, we also discuss the main outcome of the
calculations on energy spectra, electric quadrupole properties (B(E2) values, quadrupole moments), and its comparison with 
the available experimental data; in Section \ref{sec-other}, we discuss the results on
$\alpha$-hindrance factors as well as on the isotopic shifts; Section
\ref{sec-q-invariants} is devoted to the description of nuclear deformation properties of Po nuclei
as derived from the IBM-CM mean-field energy, from the study of the quadrupole shape invariants 
and from the study of the kinematic moments of
inertia, characterizing 
the yrast band structure of Po nuclei. Moreover, we compare the present results with
the nearby Pb, Hg and Pt isotopes.
Finally, in Section \ref{sec-conclu}, both the main conclusions as well as an outlook for further studies are presented.
 
\section{Experimental data: situation in the Po nuclei }
\label{sec-exp}

The even-even Po nuclei span a large region of isotopes, starting with the lightest presently know $^{190}$Po nucleus (N=106),
passing through the end of the shell, $^{210}$Po, all the way up to  N=134 at $^{218}$Po.

Many experimental complementary methods have been used to disentangle the properties over such a large interval. These nuclei
are extensively covered in the Nuclear Data Sheet reviews for A=190
\cite{Sing03}, A=192 \cite{Bagl12}, A=194 \cite{Sing06}, A=196
\cite{Xia07}, A=198 \cite{Xia02}, A=200 \cite{Kondev07},  A=202 \cite{Zhu08}, A=204 \cite{Chia10},
A=206 \cite{Kondev08}, A=208 \cite{Martin07}, and A=210 \cite{Basa14}, 
and span the region we concentrate on in the present paper. 

The experimental information (up to the end of the 90's), down to
A=192 (N=108), was obtained mainly using early fusion-evaporation 
reactions, followed by in-beam $\gamma$-ray spectroscopy, with, in particular, information on 
energy spectra systematics for A=192,194 \cite {fot97,hela96,hela99}, 
A=194 \cite{youn95}, A=196,198 \cite {alber91},
A=196,198 \cite{bern95}, A=198 \cite{lach94}, and A=198,200 \cite {maj90}.   
This particular part of the mass region has also been analyzed using $\alpha$ and 
$\beta$ decay studies of mass-separated Rn an At nuclei, resulting in information about excited 0$^+$ states in 
the isotopes with A=196-202 \cite{bijn95} and in energy spectra of the
isotopes A=200,202 \cite{bijn98}, as well as making 
use of $\alpha$-decay studies from the Rn nuclei for A=198 \cite{waut92}.
Alpha decay studies, moving from the Po nuclei into the Pb nuclei, have been a major fingerprint, 
in particular, in view of the shape coexisting structure in the daughter Pb nuclei. There exists an 
extensive set of results that has been accumulated during a period of about two decades \cite{waut93,waut94,wauters94a,bijn96,andrey99,allatt98,
andrey99a,andrey02,andrey06,vande03,vande05,duppen00,huyse01,delion14,delion13}.

It was observed that mass A=194 (see Fig.~\ref{fig-system-hg}) indicated a break in the energy scale observed for the heavier masses, 
where the energy of the 2$^+_1$ level is typically of the order of
$\sim 600-650$ keV, going down 
to an energy of about $\sim 300$ keV, therefore, dropping by a factor two.
More recenty, experimental studies spanning the 1999-2009 period have been performed at 
the JYFL at the University of Jyv\"skyl\"a using the recoil-decay-tagging (RDT) technique, studying prompt and delayed $\gamma$-rays down 
to mass A=190 \cite{vande03a, wise07}. Moreover, lifetimes
could be derived making use of the RDT technique in recoil distance Doppler-shift experiments for 
masses A=194 \cite{grahn06,grahn08}) and A=196 \cite{grahn09}. The experimental situation for the Po isotones down 
to A=190 has been summarized by Julin {\it et al}. \cite{jul01}.
Very recently, Coulomb excitation experiments have been performed using post-accelerated $^{196,198,200,202}$Po
beams at REX-ISOLDE, resulting in an important set of specific reduced E2 matrix elements, connecting the
0${^+_1}$ ground state with the excited 2$^+_{1,2}$ states (as well as for other combinations) \cite{keste15}.  
Charge radii have been measured for both the odd-A A=191-211 Po nuclei \cite{seliv13,seliv14} as well as
for the even-even A=192-210, 216, and 218 Po nuclei \cite{cocio11}, extending the ground-state information in
a major way.

\section{Theoretical approaches: the evolving situation in the Po nuclei}
\label{sec-theo}

The experimental situation, as discussed before, hints for a particular change in the structure moving down
in mass number A. The energy scale, which starts at the N=126 closed neutron shell, set by the seniority
scheme \cite{shalit63,talmi93,ressler04}, with a first excited $2^+$
state at $1181$ keV, quickly sets in a rather constant 
value of the $2^+$ energy at $\sim 650$ keV. However, over the span of a few mass units one notices (i) a quick
drop of the first excited $0^+$ from mass A=202 ($1758$ keV), down to
mass A=196 ($558$ keV), and (ii)  starting
at mass A=196, the $2^+_1$ energy (at $463$ keV) drops considerably setting into a new energy scale for the lower
masses (known only down to A=190 at present), of $\sim 250$ keV. 

In the early works of \cite{bern95,youn95} a description using a particle-core coupling model
approach was proposed, resulting from coupling the 2 proton particles
with a vibrational core system. With the experimental  
information on masses A=196, and later A=194, a systematic study was
carried out in \cite{youn97,cizew97} for
the mass region A=210 down to A=194. Here, it was pointed out that the
quick drop of, in particular, the first 
excited $0^+_2$ state, arose because of the need of using an enhanced particle-core coupling strength. This
point was later stressed by Oros {\it et al.}~\cite{oros99} who showed that a consistent set of coupling strengths
was unable to describe the observed properties, in particular, below mass A=200.

Calculations using a deformed average field \cite{may77} indicated that in studying the energy surfaces, while
at mass A=196, a soft spherical result was obtained, at mass A=192, an oblate minimum appeared, becoming the lowest 
minimum at a value of $\epsilon$ =-0.2 and an almost degenerate situation at A=190 with both oblate and prolate minima
with both $\epsilon$  values of $\sim 0.2$. Energy surface calculations, covering the $\beta-\gamma$ plane were
carried out using a deformed Woods-Saxon potential, resulting in the presence of three minima in the mass region
A=190 to A=186 \cite{oros99}. The above results were confirmed later
by Smirnova {\it et al.} \cite{smirnova03}, this time using 
a self-consistent HFB approach using the SLy4 Skyrme force \cite{bender03},
indicating rather wide minima in A=196-194, a lowest oblate minimum at A=192 and A=190, turning into prolate at 
A=188. Triggered by the new data on masses A=196 and A=194, more detailed beyond mean-field studies were carried
out \cite{grahn06,grahn08,grahn09} highlighting a detailed comparison on energy
spectra and E2 properties, indicating the interplay of an oblate and a
more spherical structure (vibrational). In their analysis, the authors
point towards a rather pure intruder character 
of the whole yrast band, including the ground state. In
our analysis (see Section \ref{sec-evolution}), we come to the conclusion that the ground state 
exhibits a rather mixed character instead of being of pure intruder nature.  
More recently, a full study of even-even nuclei 
in the Pb region \cite{yao13}, showing specific results for the Po nuclei is presented. 
It is interesting to point out that high-spin isomers have been studied in the Po too, making use of a deformed
Woods-Saxon potential, showing the effects of deformation on the specific excitation energy \cite{shi10,shi12}.

Most of the mean-field studies point towards the existence of rather
complex energy surfaces with the presence of several minima, although 
in many cases separated by small barriers. Anyhow, a common denominator is the presence of a regular configuration, 
slightly deformed (oblate or gamma unstable) or spherical, coexisting with an intruder configuration of prolate nature, corresponding
with a larger deformation as compared with the regular configuration. The lack of full calculations
in the $\beta-\gamma$ plane makes  difficult to formulate more precise  
conclusion about the particular shape of the coexisting minima.  

Within the framework of the shell-model and allowing for both the full neutron open shell, covering
N=126 to N=82, as well as allowing proton mp-mh excitations across the
Z=82 closed shell, the calculations are unfeasible nowadays. Therefore 
a truncated approach can be used starting 
from the standard IBM  allowing 
for the presence of extra pairs. This method, called IBM-CM was
proposed by Duval and Barrett \cite{duval81,duval82} and has been used in the study of shape coexistence 
in various mass region. Some early studies were 
carried out within the idea of a possible symmetry to be used within
an extended version of the IBM, including particle-hole pairs (namely
intruder I-spin \cite{hey92,hey94}), with specific applications to the Po nuclei
(references \cite{coster99,oros99}) 
in which the coupling between U(5) and SU(3) symmetries were explored and compared to the then existing data (mass
A=200 down to A=192)). 
Besides, a different symmetry, i.e., F-spin symmetry \cite{iach87}, was proposed to relate energy spectra
in nuclei with different number of protons, N$_{\pi}$,  and neutrons
pairs, N$_{\nu}$, outside of the nearest closed shells, however keeping
the sum  F=(N$_{\pi}$ +  N$_{\nu}$)/2 constant.  An application to the
Pb region was carried out by Barrett {\it et al.}~\cite{barrett91}.
The data obtained recently \cite{else11,page11,scheck11,bree14,liam14,kasia15,rapi15} from
detailed studies of the Hg nuclei on energy spectra and electric quadrupole properties (B(E2) values, Q-moments), however, 
do indicate that the more simple idea of I-spin is not holding so well. 
  
\section{The Interacting Boson Model with configuration
 mixing formalism}
\label{sec-ibm-cm}
\subsection{The formalism}
\label{sec-formalism}

The IBM-CM is an extension of the original IBM allowing to treat simultaneously
several boson configurations which correspond to different particle--hole
(p--h) shell-model excitations \cite{duval82}. 
In our case, the model space
includes the regular proton 2p configurations and a number of valence
neutrons outside of the Z=82, N=126 closed shells (corresponding to the
standard IBM treatment for the Po even-even nuclei) as well as the
proton 2h-4p configurations and the same number of valence neutrons
corresponding to a $[N]\oplus[N+2]$ boson space ($N$ being the number of
active protons, counting both proton holes and particles, plus the
number of valence neutrons outside the Z=82,N=126 closed shells,
divided by $2$ as the boson number). Consequently, the Hamiltonian for
two configuration mixing can be written as
\begin{equation}
  \hat{H}=\hat{P}^{\dag}_{N}\hat{H}^N_{\rm ecqf}\hat{P}_{N}+
  \hat{P}^{\dag}_{N+2}\left(\hat{H}^{N+2}_{\rm ecqf}+
    \Delta^{N+2}\right)\hat{P}_{N+2}\
  +\hat{V}_{\rm mix}^{N,N+2}~,
\label{eq:ibmhamiltonian}
\end{equation}
where $\hat{P}_{N}$ and $\hat{P}_{N+2}$ are projection operators onto
the $[N]$ and the $[N+2]$ boson spaces, 
respectively, $\hat{V}_{\rm mix}^{N,N+2}$  describes
the mixing between the $[N]$ and the $[N+2]$ boson subspaces, and
\begin{equation}
  \hat{H}^i_{\rm ecqf}=\varepsilon_i \hat{n}_d+\kappa'_i
  \hat{L}\cdot\hat{L}+
  \kappa_i
  \hat{Q}(\chi_i)\cdot\hat{Q}(\chi_i), \label{eq:cqfhamiltonian}
\end{equation}
is a restricted IBM Hamiltonian called  extended consistent-Q Hamiltonian (ECQF) \cite{warner83,lipas85} with $i=N,N+2$,
$\hat{n}_d$ the $d$ boson number operator, 
\begin{equation}
  \hat{L}_\mu=[d^\dag\times\tilde{d}]^{(1)}_\mu ,
\label{eq:loperator}
\end{equation}
the angular momentum operator, and
\begin{equation}
  \hat{Q}_\mu(\chi_i)=[s^\dag\times\tilde{d}+ d^\dag\times
  s]^{(2)}_\mu+\chi_i[d^\dag\times\tilde{d}]^{(2)}_\mu~,
\label{eq:quadrupoleop}
\end{equation}
the quadrupole operator. 
This
approach has been proven  to be a good approximation in several recent
papers on Pt \cite{Garc09,Garc11} and Hg isotopes \cite{Garc14}. 

The parameter $\Delta^{N+2}$ can be
associated with the energy needed to excite two proton particles across the
Z=82 shell gap, giving rise to 2p-2h excitations, corrected for the pairing interaction gain and including
monopole effects~\cite{hey85,hey87}.
The operator $\hat{V}_{\rm mix}^{N,N+2}$ describes the mixing between
the $N$ and the $N+2$ configurations and is defined as
\begin{equation}
  \hat{V}_{\rm mix}^{N,N+2}=w_0^{N,N+2}(s^\dag\times s^\dag + s\times
  s)+w_2^{N,N+2} (d^\dag\times d^\dag+\tilde{d}\times \tilde{d})^{(0)}.
\label{eq:vmix}
\end{equation}

The E2 transition operator for two-configuration mixing is
subsequently defined as
\begin{equation}
  \hat{T}(E2)_\mu=\sum_{i=N,N+2} e_i
  \hat{P}_i^\dag\hat{Q}_\mu(\chi_i)\hat{P}_i~,\label{eq:e2operator}
\end{equation}
where the $e_i$ ($i=N,N+2$) are the effective boson charges and
$\hat{Q}_\mu(\chi_i)$ the quadrupole operator defined in equation
(\ref{eq:quadrupoleop}).

In section \ref{sec-fit-procedure} we present the methods used in order to 
determine the parameters
appearing in the IBM-CM Hamiltonian as well as in the $\hat{T}(E2)$ operator.

The wave function, within the IBM-CM, can be described as 
\begin{eqnarray}
\Psi(k,JM) &=& \sum_{i} a^{k}_i(J;N) \psi((sd)^{N}_{i};JM) 
\nonumber\\
&+& \
\sum_{j} b^{k}_j(J;N+2)\psi((sd)^{N+2}_{j};JM)~,
\label{eq:wf:U5}
\end{eqnarray}
where $k$, $i$, and $j$ are rank numbers. The  weight of the wave function contained within the 
$[N]$-boson subspace, can then be defined as
the sum of the squared amplitudes $w^k(J,N) \equiv \sum_{i}\mid a^{k}_i(J;N)\mid ^2$. Likewise,
one obtains the content in the $[N+2]$-boson subspace.

\subsection{The fitting procedure: energy spectra and absolute B(E2) reduced 
transition probabilities}
\label{sec-fit-procedure}

Here, we present the way in which the parameters of the Hamiltonian
(\ref{eq:ibmhamiltonian}), (\ref{eq:cqfhamiltonian}),
(\ref{eq:quadrupoleop}),  and (\ref{eq:vmix})  
and the effective charges in the $\hat{T}(E2)$ transition operator
(\ref{eq:e2operator}) have been determined. 
We study the range $^{190}$Po to $^{208}$Po, thereby, covering
almost the whole second half of the neutron shell N=82-126. 
\begin{table}[hbt]
\caption{Energy levels, characterized by $J^{\pi}_i$, included in
  the energy fit, if known, and the assigned $\sigma$ values in
  keV.}
\label{tab-energ-fit}
\begin{center}
  \begin{tabular}{|c|c|}
    \hline
    Error (keV) & States  \\
    \hline
    $\sigma=0.1$& $2_1^+$ \\
    $\sigma=1$ & $4_1^+, 0_2^+, 2_2^+$\\
    $\sigma=10$  & $2_3^+, 3_1^+, 4_2^+, 6_1^+, 8_1^+$ \\
    $\sigma=100$   & $6_2^+$\\
    \hline
  \end{tabular}
\end{center}
\end{table}

In the fitting procedure carried out here, we try to obtain the best possible agreement
with the experimental data including both the excitation energies
and the B(E2) reduced transition probabilities. 
Using the expression of the IBM-CM Hamiltonian, as given in equation (\ref{eq:ibmhamiltonian}),
and of the E2 operator, as given in equation (\ref{eq:e2operator}), in the most general case $13$
parameters show up.  
We impose as a constraint to obtain parameters that change smoothly  
in passing from isotope to isotope.  Note also that we constrained 
$\chi_{N}=0$ and $\kappa'_{N+2}=0$. We have explored in detail the validity of this
assumption and we have found very little improvement in the value of
$\chi^2$ (see Eq.~(\ref{chi2})) when releasing those parameters.   
On the other hand, we have kept the value that describes the
energy needed to create an extra particle-hole pair ($2$ extra bosons) constant, 
{\it i.e.}, $\Delta^{N+2}=2400$ keV, and have also put the constraint of keeping the
mixing strengths constant too, {\it i.e.}, $w_0^{N,N+2}=w_2^{N,N+2}=30$ keV for all the Po isotopes. 
We also have to determine for each isotope the effective charges of the $E2$ operator.
This finally leads to $8$ parameters to be varied in each nucleus. 

\begin{figure}[hbt]
  \centering
  \includegraphics[width=0.5\linewidth]{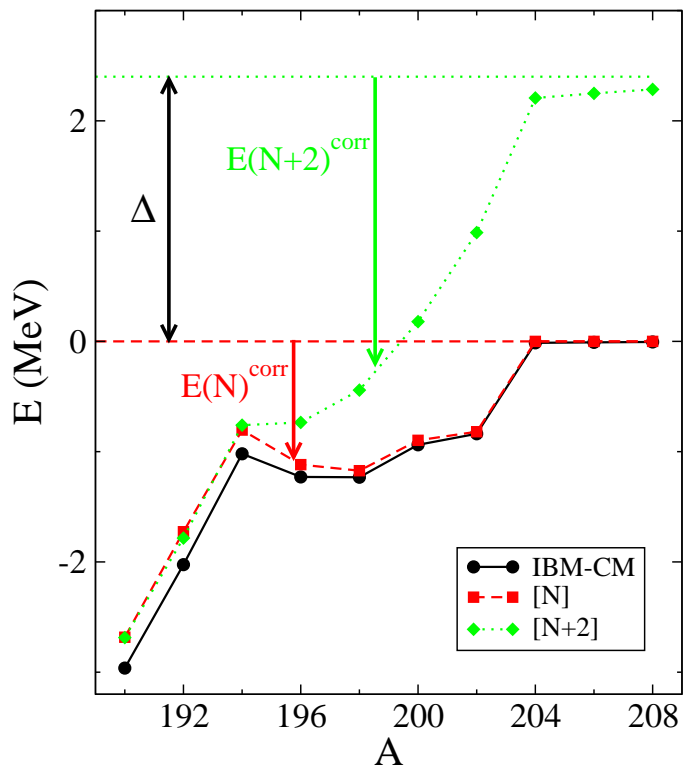}
  \caption{(Color online) Absolute energy of the lowest unperturbed regular and
    intruder 0$^+_1$ states 
    for $^{190-208}$Po. The arrows correspond to the correlation
    energies in the N and N+2 subspaces (see also the text for a more detailed discussion).}
  \label{fig-energ-corr}
\end{figure}

To determine the value of $\Delta^{N+2}$ we considered as a reference the already
known values for Pt and Hg, which are $2800$ keV and $3480$ keV,
respectively. On one hand, we expect to get a value closer to Pt than
to Hg because the intruder states for Po are supposed to be very close
to the regular ones and even become the ground state around the mid-shell. That constrains
our value to be not much higher than $2800$ keV. On the other hand, we
know that the heavier isotopes, near the closed shell,
should have an energy gap between the regular and the intruder states which
is equal, at maximum, to $\Delta^{N+2}$ (see Fig.~\ref{fig-energ-corr} on correlation
energy) and, moreover, experimentally the intruder states should appear above $2000-3000$ keV.     

To determine the value of the mixing strengths, we considered that the
corresponding value for the Pt nuclei was fixed to $50$ keV \cite{Garc09}, while for
the Pb, to a smaller strength of $18$ keV \cite{fossion03,helle05}, and
for Hg  to $20$ keV. We performed a set
of exploratory calculations between $20-30$ keV and found that the best overall agreement corresponds
to $w_0^{N,N+2}=w_2^{N,N+2}=30$ keV, although the agreement is quite
similar over the whole range.
\begin{table}[hbt]
\caption{Hamiltonian and $\hat{T}(E2)$ parameters resulting from the present study.
 All quantities have the dimension of energy (given in units of keV),
except $\chi_{N+2}$ which is dimensionless and $e_{N}$ and $e_{N+2}$
which are given in units $\sqrt{\mbox{W.u.}}$ 
The remaining parameters of the Hamiltonian, {\it i.e.}, $\chi_N$,
and $\kappa'_{N+2}$ are equal to zero, except for $\Delta^{N+2}=2400$ keV
and $w_0^{N,N+2}=w_2^{N,N+2}=30$ keV. }
\label{tab-fit-par-mix}
\begin{center}
\begin{ruledtabular}
\begin{tabular}{ccccccccc}
Nucleus&$\varepsilon_N$&$\kappa_N$&$\kappa'_N$&$\varepsilon_{N+2}$&
$\kappa_{N+2}$&$\chi_{N+2}$&$e_{N}$&$e_{N+2}$\\
\hline
$^{190}$Po&712.6& -29.41&   6.34& 285.43& -28.56& 0.22 & 2.88\footnotemark[1]& 1.86\footnotemark[1]\\ 
$^{192}$Po&731.8& -24.20&   0.43& 402.67& -30.09& 0.21 & 2.88\footnotemark[1]& 1.86\footnotemark[1]\\ 
$^{194}$Po&800.3& -15.12& -16.00& 518.23& -28.92& 0.41 & 2.88& 1.86\\ 
$^{196}$Po&845.0& -22.76& -10.96& 373.16& -31.02& 0.15 & 1.86& 1.86\\ 
$^{198}$Po&982.3& -28.00& -25.74& 854.82& -34.51& 1.33 & 2.05& 1.10\\ 
$^{200}$Po&955.8& -26.05& -25.00& 843.48& -34.55& 1.31 & 2.01& 1.10\footnotemark[2]\\
$^{202}$Po&942.1& -29.06& -28.46& 236.30& -20.63& 1.09 & 2.28& 1.10\footnotemark[2]\\
$^{204}$Po&810.4&  -0.00& -21.09& 100.00&  -5.00& 0.50 & 2.28\footnotemark[3]& 1.10\footnotemark[2]\\
$^{206}$Po&717.1&  -0.00&  -4.30& 100.00&  -5.00& 0.50 & 2.28\footnotemark[3]& 1.10\footnotemark[2]\\ 
$^{208}$Po&643.4&  -0.00&   6.52& 100.00&  -5.00& 0.50 & 2.28\footnotemark[3]& 1.10\footnotemark[2]\\
\end{tabular}
\end{ruledtabular}
\end{center}
\footnotetext[1]{The effective charges have been taken the same as the corresponding 
  values obtained for $^{194}$Po.}
\footnotetext[2]{$e_{N+2}$ corresponding to $^{198}$Po.}
\footnotetext[3]{$e_{N}$ corresponding to  $^{202}$Po.}
\end{table}

The $\chi^2$ test is used in the fitting procedure in order to extract the optimal solution. 
The $\chi^2$ function is defined in the standard way as
\begin{equation}
  \label{chi2}
  \chi^2=\frac{1}{N_{data}-N_{par}}\sum_{i=1}^{N_{data}}\frac{(X_i
    (data)-X_i (IBM))^2}{\sigma_i^2},
\end{equation} 
where $N_{data}$ is the number of experimental data,
$N_{par}$ is the number of parameters used in the IBM fit, $X_i(data)$
describes the experimental excitation energy of a given excited state (or an experimental 
B(E2) value), $X_i(IBM)$ denotes the corresponding calculated IBM-CM value,
and $\sigma_i$ is an error (theoretical) assigned to each $X_i(data)$ point. 
We minimize the $\chi^2$
function for each isotope separately using the package MINUIT \cite{minuit} which allows to
minimize any multi-variable function.

In some of the lighter Po isotopes, due to the small number of experimental data, the
values of some of the free 
parameters could not be fixed unambiguously using the above fitting
procedure. Moreover, for the heavier isotopes (A$>$202), that part of the
Hamiltonian corresponding to the intruder states is fixed such as to guarantee 
that those states appear well above the regular ones, that
is, above $2$ MeV. In some cases due to the lack of experimental data
the effective charges could not be determined.

As input values, we have used the excitation energies of the levels presented in 
Table \ref{tab-energ-fit}. In this table we also give the corresponding $\sigma$ values. 
We stress
that the $\sigma$ values do not correspond
to experimental error bars, but they are related with the expected
accuracy of the IBM-CM calculation to reproduce a particular experimental data point. 
Thus, they act as a guide so that a given calculated level converges towards the
corresponding experimental level.
The $\sigma$ ($0.1$ keV) value for the $2_1^+$ state guarantees
the exact reproduction of this experimental most important excitation energy, {\it
  i.e.}, the whole energy spectrum is normalized to this experimental
energy. The states $4_1^+, 0_2^+$ and $2_2^+$ are 
considered as the most important ones to be reproduced ($\sigma=1$
keV). The group of states  
$2_3^+$, $3_1^+$, $4_2^+$, $6_1^+$, and $8_1^+$ ($\sigma=10$
keV) and  
$6_2^+$ ($\sigma=100$ keV) should also be well
reproduced by the calculation to guarantee a correct moment of
inertia for the yrast band and the structure of the $0_2^+$ band. Note
that we only considered states in the fit with angular momentum and parity
unambiguously determined.

In the case of the $E2$ transitions rates, we have used the available
experimental data involving the states presented in Table \ref{tab-energ-fit},
restricted to those $E2$ transitions for which absolute B(E2) values
are known, except if serious hints that the states involved 
present non-collective degrees of freedom exist. 
Additionally, we have taken a value of $\sigma$ that corresponds to $10\%$ of the 
B(E2) values or to the experimental error bar if larger, except
for the transition $2_1^+\rightarrow 0_1^+$ where a smaller value of
$\sigma$ ($0.1$ W.u.) was taken, thereby normalizing, in most of cases, our calculated values to the
experimental $B(E2;2^+_1 \rightarrow 0^+_1)$ value. 

This has resulted in the values of the parameters for the IBM-CM Hamiltonian,
as given in Table \ref{tab-fit-par-mix}. 
In the case of $^{190-192}$Po and $^{200-208}$Po, 
the value of the effective charges, or part of them, cannot be determined
because not a single absolute B(E2) value is known or $\chi^2$ is
insensitive to their values. However, for
completeness we have taken the effective charges of $^{194}$Po for $^{190-192}$Po, $e_{N+2}$ of
$^{198}$Po for $^{200-208}$Po, and $e_{N}$ of $^{202}$Po for $^{204-208}$Po.

\subsection{Correlation energy in the configuration mixing approach} 
\label{sec-corr_energy}

Intruder states are expected to appear, in principle, at an excitation
energy well above the corresponding regular ones with identical
angular momentum. The reason is that these configurations are related
to the creation of a 2p-2h excitation across the Z=82 closed shell. In
the case of Po this energy would correspond to $\Delta^{N+2}=2400$
keV. However, according to reference \cite{hey87}, the single-particle
energy cost has to be corrected because of the strong pairing energy
gain when forming two extra $0^+$ coupled (particle and hole) pairs,
by the quadrupole energy gain when opening up the proton shell, as well
as by the monopole correction caused by a change in the
single-particle energy gap at Z=82 as a function of the neutron
number. In particular, around the midshell point at N=104, where the number of
active nucleons becomes maximal, the energy gain due to the
strong correlation energy is such that the energy of the intruder configurations 
becomes close to the energy of the regular ones. It may even be that a crossing results
making the intruder state forming a ground state. 
For instance, in the case of Pt isotopes, the nuclei
around the midshell point at N=104 exhibit a ground state of intruder nature
\cite{Garc09,Garc11}. 
\begin{figure}[hbt]
  \centering
  \includegraphics[width=0.8\linewidth]{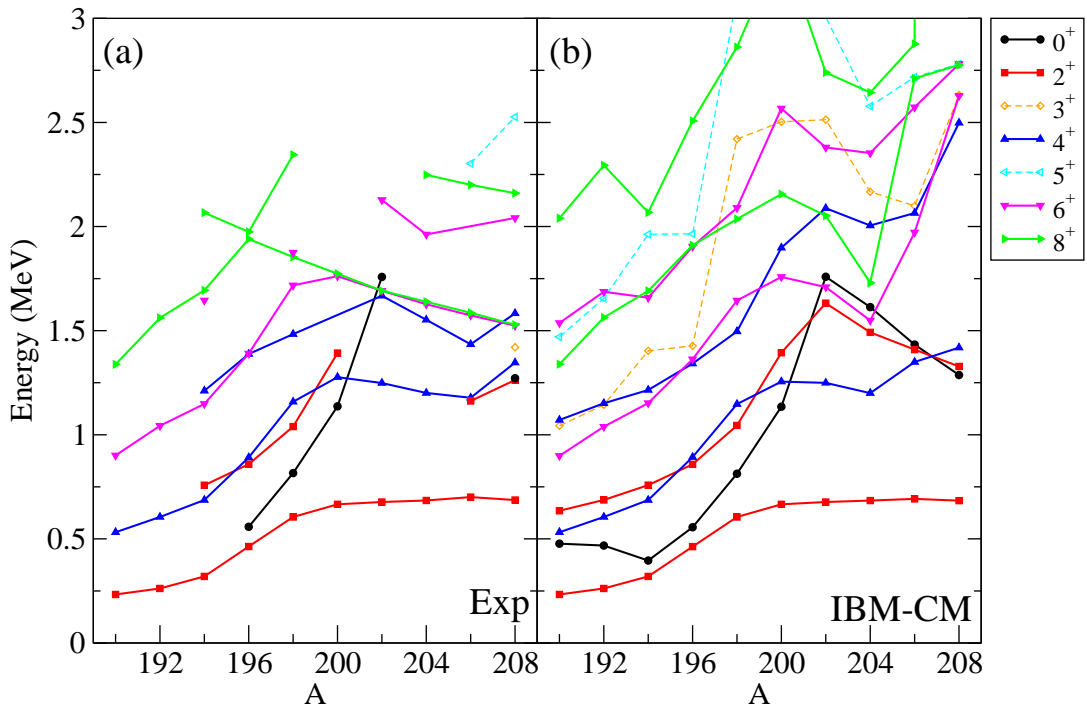}
  \caption{(Color online) Experimental excitation energies (up to $E_x \approx 3.0$
    MeV) (panel (a)) and the theoretical results (panel (b)),
    obtained from the IBM-CM.} 
  \label{fig-energ-comp}
\end{figure}

The experimental states can present a strong mixing because of the 
interaction between both families of configurations (regular and intruder). Therefore, it is not simple
in many cases to find out which configurations are dominant in the ground-state wavefunction. 
In order to take advantage of the IBM-CM calculations, we calculate
explicitly the ``absolute'' energy of the lowest $0^+$ state belonging
to both the regular [N] and [N+2] intruder configuration spaces, turning off the
interaction among the two families. We choose as the reference energy, the
energy of the regular Hamiltonian describing the [N] space for the $^{210}$Po nucleus. This nucleus
only consists of a single boson, with the s-boson as the lowest energy state, resulting in the
zero-energy line (horizontal dashed red line in Fig.~\ref{fig-energ-corr}).
The energy of the lowest 0$^+$ state in the regular configuration space [N], E(0$^+_1$,N), is lowered with respect to
the reference energy because of the correlation energy and is described by the wave function
$\Psi(0^+_1)^{reg}_N$  (see also expression (\ref{eq:wf:N})). The lowering
depends on the number of bosons. On the other hand, 
the energy of the lowest 0$^+$ state in the intruder configuration space [N+2], E(0$^+_1$,N+2), is described by
the wave function $\Psi(0^+_1)^{int}_{N+2}$ (see also expression (\ref{eq:wf:N+2})) and appears at the 
energy corresponding to
$\Delta^{N+2}$. This energy will subsequently be lowered by its specific correlation energy too. 
In most cases, the regular configuration with N bosons corresponds to a
spherical or slightly deformed shape, while the intruder ones, with N+2
bosons, to a more deformed shape. Therefore, the energy gain for the
lowest intruder state, described by the wave function $\Psi(0^+_1)^{int}_{N+2}$, uses to be larger than for the lowest regular 
state, described by the wave function $\Psi(0^+_1)^{reg}_N$. The
relative position of these lowest regular and intruder states is
plotted in Fig.~\ref{fig-energ-corr}. Here, it can be clearly appreciated how the
energies of both configurations can become very close, depending on the balance between the
off-set, $\Delta^{N+2}$, and the difference in the correlation energy $E(N+2)^{corr} - E(N)^{corr}$.    

One notices how the energies become really close near
midshell (N=106, A=190), where the number of active nucleons is maximal, showing that 
the lowest intruder state can determine the character of the lowest 0$^+_1$ state. From A=196,
and moving towards the heavier masses, 
the energy difference starts to increase 
and both states appear to be very well separated. Note that
once the isotopes approach the end of the shell, at N=126, the number of active
bosons is drastically reduced and, therefore, the correlation energy
is reduced dramatically. In particular, the energy of the regular
state reaches the 
reference energy, which means that it corresponds to essentially a spherical shape, 
while the energy of the intruder state approaches the energy
$\Delta^{N+2}$ quite closely. Consequently, at the end (or at the beginning) of the
shell, the maximum energy difference between both states will
correspond to $\Delta^{N+2}$. The existence of a ceiling for the
energy difference modulates the parabolic behavior of the energy
systematics of the intruder states, transforming it into a flat shape
and, therefore, leading to an energy for the intruder states lower
than experimentally observed. Clearly, this is a deficiency of the IBM-CM
calculations near the shell closure.   

\begin{figure}[hbt]
  \centering
  \includegraphics[width=0.6\linewidth]{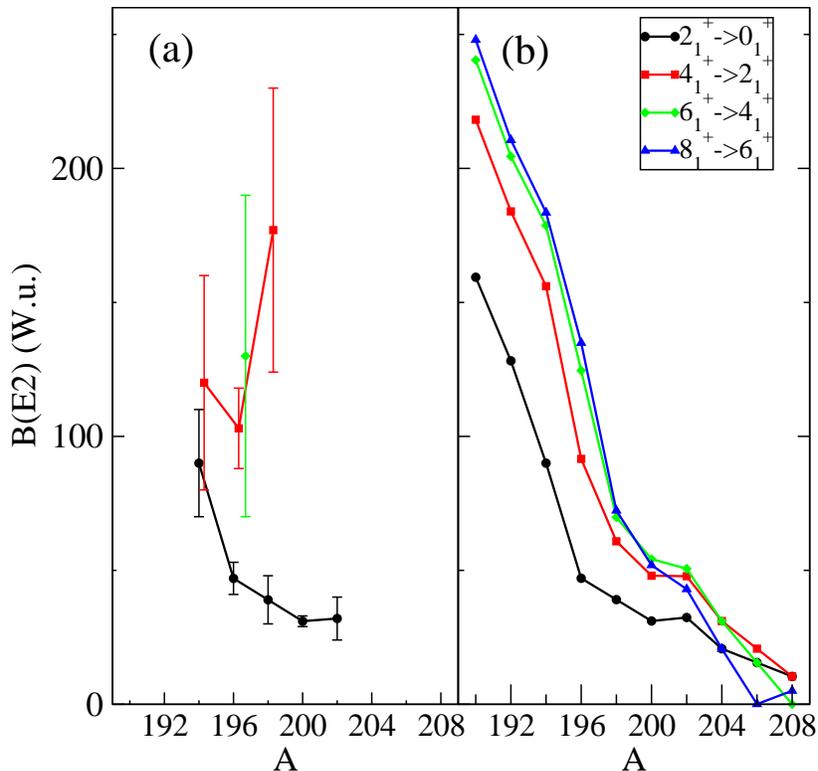}
  \caption{(Color online) Comparison of the absolute B(E2) reduced transition
    probabilities along the yrast band, given in W.u. Panel (a) corresponds to known
    experimental data and panel (b) to the theoretical IBM-CM
    results.} 
  \label{fig-be2-1}
\end{figure}

\begin{figure}[hbt]
  \centering
  \includegraphics[width=0.6\linewidth]{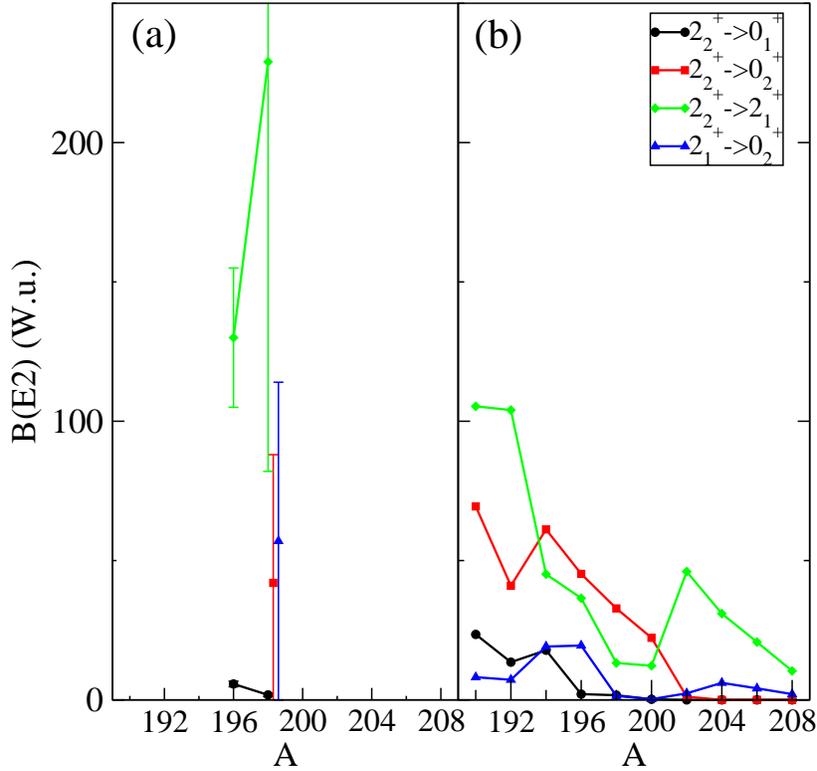}
  \caption{(Color online) Comparison of the few non-yrast intraband
    absolute B(E2) reduced transition 
    probabilities, given in W.u. Panel (a) corresponds to the few known
    experimental data, panel (b) to the theoretical IBM-CM results.}
  \label{fig-be2-2}
\end{figure}

\begin{table}
  \caption{Comparison of the experimental absolute B(E2) values (given in
    units of W.u.) with
    the IBM-CM Hamiltonian results.
    Data are taken from the Nuclear Data
    Sheets~\cite{Sing03,Bagl12,Sing06,Xia07,Xia02,Kondev07},  
    complemented with references presented in section
    \ref{sec-exp}.}  
  \label{tab-be2a}
\begin{center}
\begin{ruledtabular}
\begin{tabular}{cccc}
Isotope   &Transition             &Experiment&IBM-CM \\
\hline
$^{194}$Po&$2_1^+\rightarrow 0_1^+$& 90(20)         & 90    \\  
         &$4_1^+\rightarrow 2_1^+$& 120(40)       & 156 \\   
\hline
$^{196}$Po&$2_1^+\rightarrow 0_1^+$& 47(6)         & 47    \\  
         &$4_1^+\rightarrow 2_1^+$& 103(15)       & 92 \\   
         &$6_1^+\rightarrow 4_1^+$& 130(60)       & 125   \\   
         &$2_2^+\rightarrow 0_1^+$& 5.7(10)\footnotemark[1]      &  2.1  \\   
         &$2_2^+\rightarrow 2_1^+$& 130(25)\footnotemark[1]       & 36   \\   
\hline
$^{198}$Po&$2_1^+\rightarrow 0_1^+$& 39(9)         & 39    \\  
         &$4_1^+\rightarrow 2_1^+$& 177(53)       & 61 \\   
         &$8_1^+\rightarrow 6_1^+$& 2.0(1)\footnotemark[2]$^,$ \footnotemark[3] & 72 \\   
         &$0_2^+\rightarrow 2_1^+$& 285$(^{+980}_{-285})$\footnotemark[1]  &  7  \\   
         &$2_2^+\rightarrow 0_1^+$& 1.8$(^{+1.6}_{-0.6})$\footnotemark[1]  &  1.7  \\   
         &$2_2^+\rightarrow 2_1^+$& 229(147)\footnotemark[1]      & 13   \\   
         &$2_2^+\rightarrow 0_2^+$& 42(56)\footnotemark[1]      & 33   \\   
\hline
$^{200}$Po&$2_1^+\rightarrow 0_1^+$& 31(2)         & 31    \\  
         &$8_1^+\rightarrow 6_1^+$& 9.4(5)\footnotemark[2] &  52\\   
\hline
$^{202}$Po&$2_1^+\rightarrow 0_1^+$& 32(8)         & 32    \\  
\end{tabular}
\end{ruledtabular}
\end{center}
\footnotetext[1]{Data taken from Ref.~\cite{keste15}.}
\footnotetext[2]{Experimental data not included in the fit.}
\footnotetext[3]{Data taken from Ref.~\cite{maj90}.}
\end{table}

\begin{figure}[hbt]
  \centering
  \includegraphics[width=0.5\linewidth]{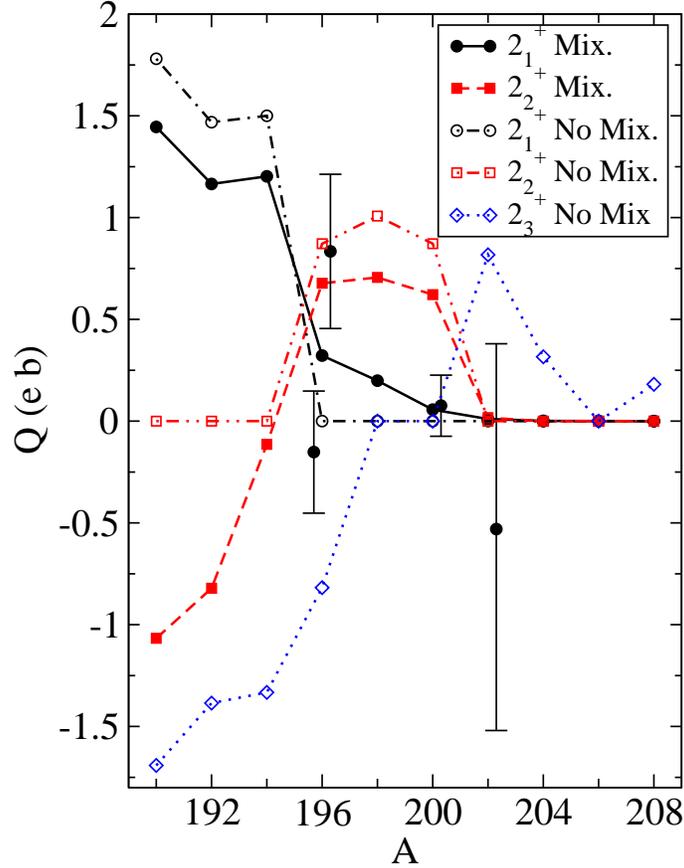}
  \caption{(Color online) IBM-CM values of the quadrupole moments for the 
    $2_1^+$ and $2_2^+$ states in $^{190-208}$Po. The quadrupole moments are given
    in units $e\, b$. The dash-dotted and dotted lines indicate the quadrupole moments when the mixing Hamiltonian is
    switched off, corresponding to the unperturbed $2_1^+$, $2_2^+$, and $2_3^+$ states.}
  \label{fig-q-theo}
\end{figure}

\subsection{Detailed comparison between the experimental data and the IBM-CM results:
 energy spectra and electric quadrupole properties}
\label{sec-fit-compare}

In this subsection, we compare in detail the experimental
energy spectra with the theoretical ones up to an excitation energy
below $E_x\approx 3.0$ MeV. In Fig.~\ref{fig-energ-comp}a we plot the
experimental data set, while in  Fig.~\ref{fig-energ-comp}b the
theoretical values are depicted. Note that the energy
corresponding to the $2_1^+$ state perfectly agrees with the
theoretical one because this level was used to normalize the energies,
in other words, we used in the fitting procedure for this level a very small value
for $\sigma$ ($0.1$ keV) in order to exactly reproduce the $2_1^+$ experimental
energy. Comparing the energy spectra, one can distinguish two regions, $A<200$ and
$A\geq 200$, with an overall better agreement in the first region as compared to the
second. A most probable reason for this is the fact that in the
lightest isotopes the 
number of active bosons is substantially larger and the collectivity is enhanced. As
we discuss later, in the heavier Po isotopes, non-collective broken-pair states
(with a v=2 seniority character) 
do appear in the low-energy part of the energy spectrum. Those states
are not well reproduced within the IBM space because of the restriction to a set
of interacting s and d bosons. 

One observes in Fig.~\ref{fig-energ-comp} how the compression of the
energy spectrum, with decreasing mass number, 
is well reproduced. The same happens for the systematic dropping of the
energy of the $0_2^+$ states, and the energy of the yrast band, more generally. 
Note that the agreement is better for those states with low and even angular momenta.

Next, we carry out a comparison for the B(E2) values, which is a much
more stringent test than the excitation energy, because these numbers are highly
dependent on the detailed structure of the wave function. Although the
existing experimental information on B(E2) values is rather scarce,
recent new results from Coulomb excitation
experiments at REX-ISOLDE, in particular for the $^{196-198}$Po isotopes, have
appeared \cite{keste15}, improving the experimental knowledge of this mass region.  

In Figs.~\ref{fig-be2-1} and \ref{fig-be2-2} we compare the B(E2) reduced transition
probabilities, while in
Fig.~\ref{fig-q-theo} we compare the electric quadrupole moments. We
also present a more detailed comparison on B(E2) values in Table \ref{tab-be2a}. 

In Fig.~\ref{fig-be2-1} we present the intraband B(E2) values along the 
yrast band, for which the most complete experimental information
exists. In a similar way as for the excitation energies, we used the
$B(E2; 2_1^+\rightarrow 0_1^+)$ reduced transition probability to normalize the
theoretical results. Note that for some nuclei, where 
experimental information is scarce, we have used the corresponding values of
the neighboring  isotopes to determine the effective charges (as explained
in Sec.~\ref{sec-fit-procedure}). On the experimental side, the value
of $B(E2; 2_1^+\rightarrow 0_1^+)$ is systematically dropping, which
denotes a reduction of the collectivity of the states when
approaching the end of the neutron shell. However, this is not the case for
$B(E2; 4_1^+\rightarrow 2_1^+)$ in $^{198}$Po, where an unexpected large
value is observed. On the other hand, the theoretical results show a continous drop of
all the intraband transitions, resulting from the reduction of
the collectivity of the states (in line with a decreasing number of active bosons). 

In Fig.~\ref{fig-be2-2} we study the systematics of some interband
transitions involving  the states $0_{1,2}^+$ and
$2_{1,2}^+$. Because of the lack of experimental data and the large
error bars, it is difficult to extract any trend. Regarding the
theoretical results, once more, one notices an overall reduction of
the B(E2) values. We point out that for $A>196$, the
observed B(E2) values are of the same order as the observed B(E2) values in the yrast
band. This represents a hint about the changing mixing character
of the states as a function of mass number A (see
Fig.~\ref{fig-overlap} as illustration of the changing structure of
the wave function).  

A different way to extract information on the changing character of the wave functions 
for the lowest two $2^+$ states is comparing the spectroscopic quadrupole moments, shown in
Fig.~\ref{fig-q-theo}. Here, we compare the theoretical
and the experimental values of the quadrupole moment for the states
$2_1^+$ and $2_2^+$. Moreover, we present the values corresponding to
the unperturbed lowest-lying three $2^+$ states. One notices that the $2^+_1$ state
corresponds to a well deformed and oblate shape for the lightest isotopes, which is 
smoothly changing into a rather spherical shape for A=202 and onwards. On
the other hand, the $2^+_2$ state corresponds to a prolate shape for the lightest isotopes,
changing into an oblate shape for A=196-200, and into a spherical
shape for A=202 and the still heavier isotopes. When comparing with the unperturbed
values, one can see how the $2_1^+$ state is built up as 
a mixture of the first two unperturbed $2^+$ states. However, in order to understand the 
second $2_2^+$ state, we have to resort to a mixture of  
the second and third unperturbed $2^+$ states mainly. This figure presents a clear proof of the changing character for
the first two $2^+$ states as a function of mass number. This issue will be discussed in section
\ref{sec-evolution} in a more quantitative way. 

\begin{figure}[hbt]
  \centering
  \includegraphics[width=1\linewidth]{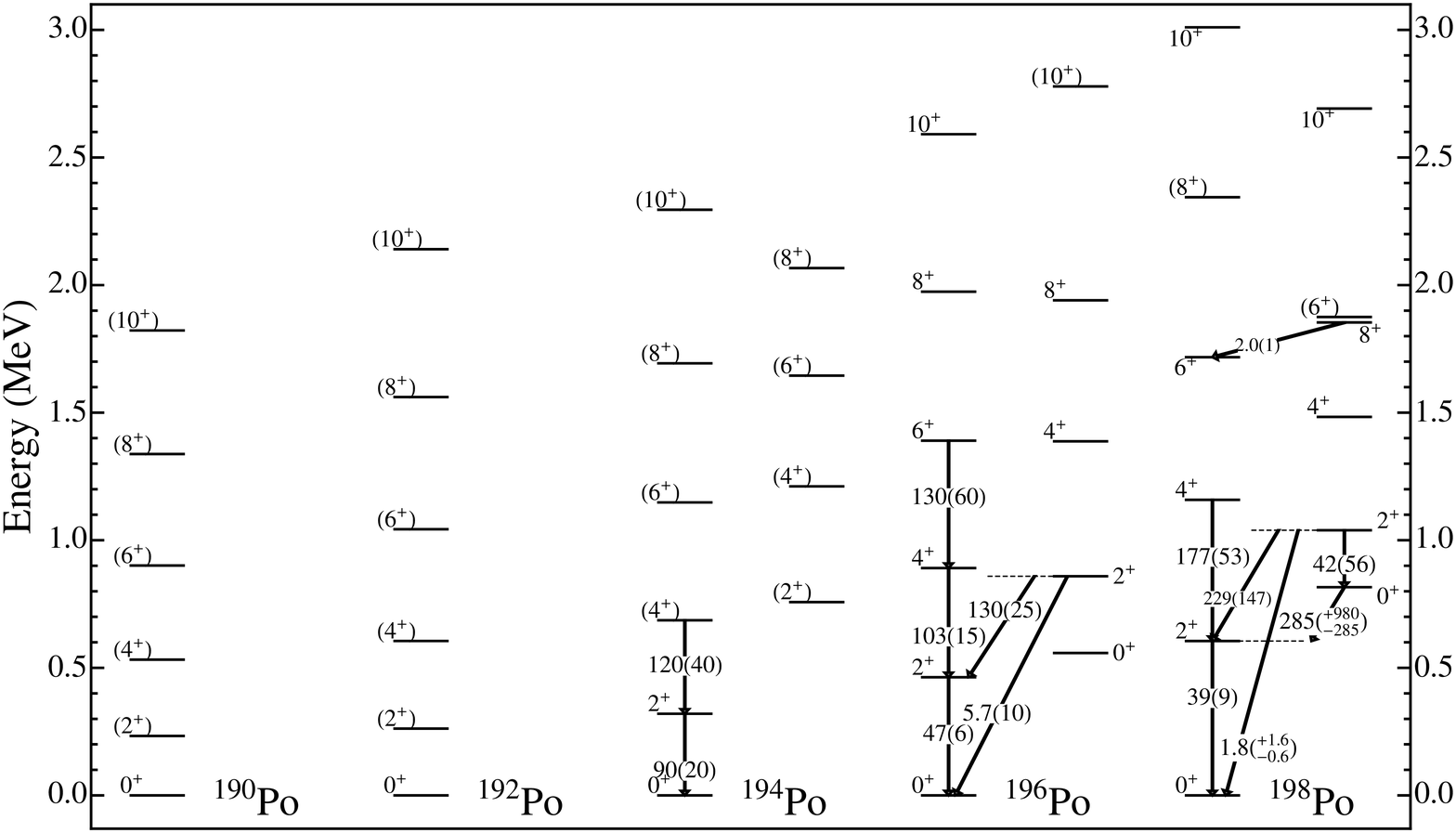}
  \caption{Experimental excitation energies and absolute B(E2)
    transition rates for selected
    states in $^{190-198}$Po.} 
  \label{fig-exp-190-198}
\end{figure}

\begin{figure}[hbt]
  \centering
  \includegraphics[width=1\linewidth]{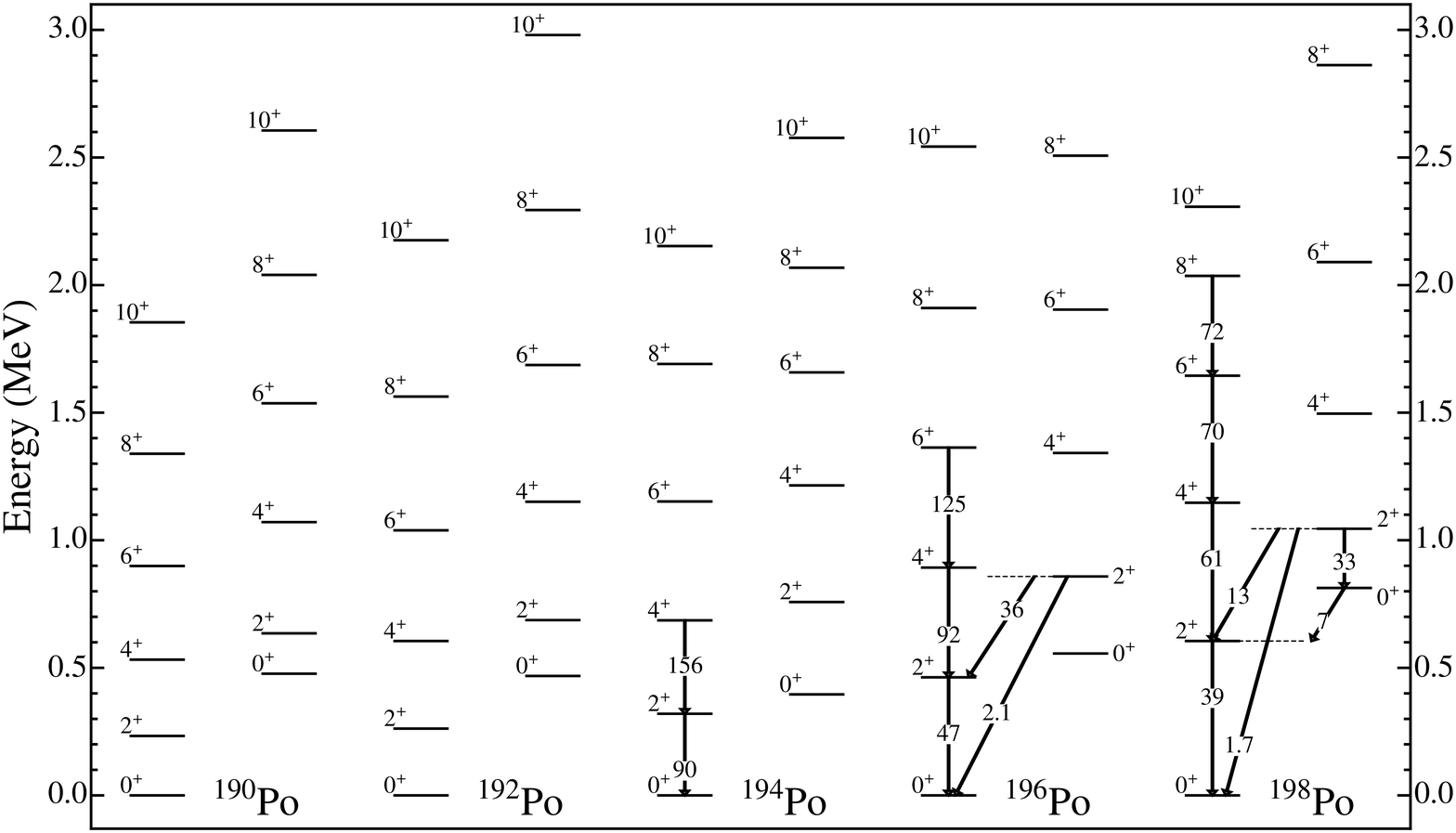}
  \caption{Theoretical excitation energies and absolute B(E2)
    transition rates for selected states in $^{190-198}$Po.} 
  \label{fig-theo-190-198}
\end{figure}

In Figs.~\ref{fig-exp-190-198} and \ref{fig-theo-190-198} we present
the experimental and theoretical energy spectra (up to $E_x\approx 2.5$ MeV) 
for masses A=190-198, which is 
the region where the coexistence should be more evident. We include in
the comparison the known absolute B(E2) values. One observes a rather distinct rotational 
structure (along the yrast band) for A=190-194,
followed by a structure that is changing into a more vibrational behavior for masses A=196-198. 

\begin{figure}[hbt]
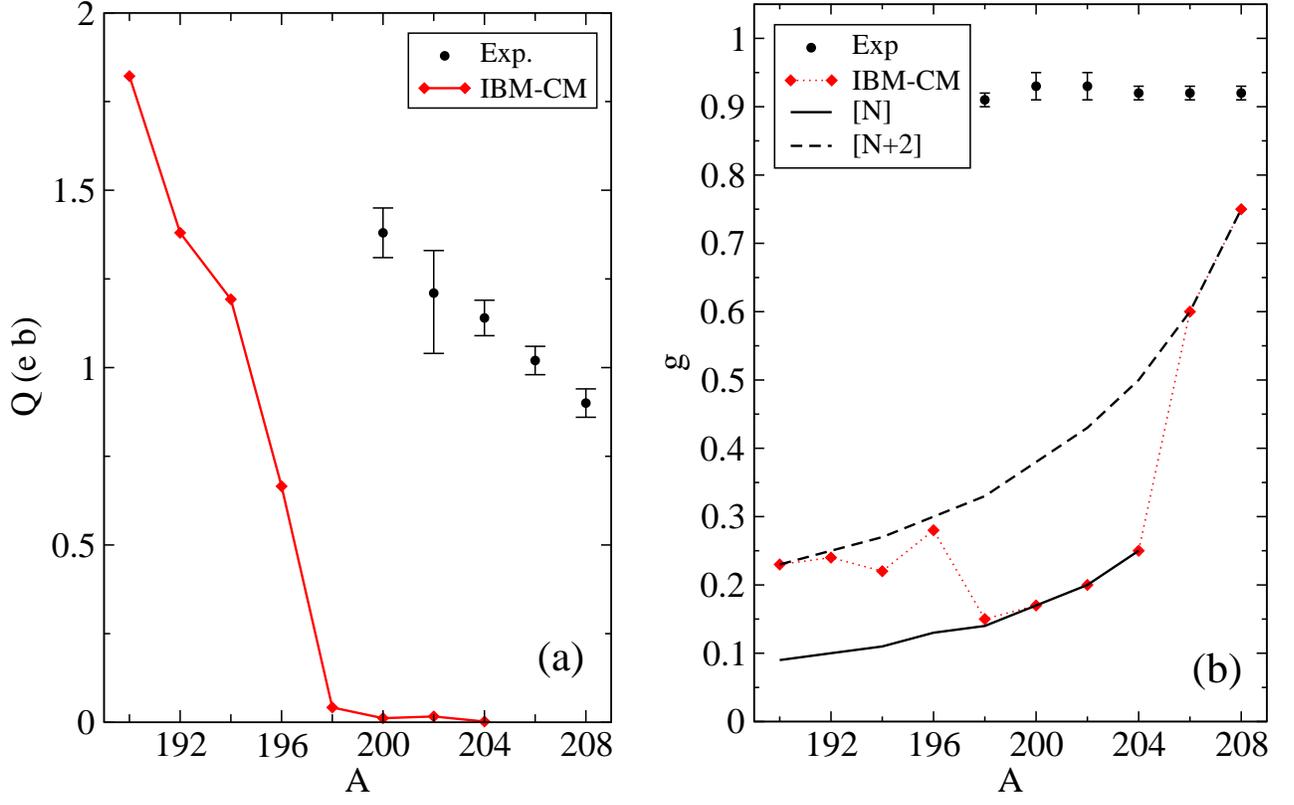

  \centering
  \includegraphics[width=.45\linewidth]{Quadrupole-8.eps}
~~~~
  \includegraphics[width=.45\linewidth]{g-8_1.eps}%
  \caption{(Color online) Panel (a): Experimental and theoretical quadrupole moment for the $8_1^+$
    state (note that the theoretical values for A=206 and 208 are not
    shown due to the reduced number of valence bosons). Panel (b): Gyromagnetic
    factor for the $8_1^+$ state (experimental data and theoretical results).}
  \label{fig-state8-1}
\end{figure}

The Po energy spectra show some other most interesting structural changes starting at the neutron
N=126 shell closure, moving towards the neutron-deficient region (decreasing towards mid-shell
at  A=188) (see Figs.~\ref{fig-system-hg} and \ref{fig-energ-comp}). One starts with the typical $\pi(1h_{9/2})^2$ two-particle spectrum governed
by the seniority characteristics of the two-body proton-proton interaction.  An energy gap of
$\sim$ 0.8 MeV is observed from the 8$^+_1$ to the 8$ ^+_2$ state
energy difference (see Fig.~\ref{fig-system-hg}), 
which most probably results from 
a $\pi(1h_{9/2}2f_{7/2})$ configuration, as a measure of the single-particle energy gap.
Next, one observes the remaining characteristics of a senority v=2 spectrum, albeit with an immediate
drop of the 2$^+_1$ state, when removing two neutrons from the closed N=126 shell, which then
keeps a remarkably constant energy down to mass A=200 (N=116). The energy of the 4$ ^+_1$ slighty
drops but also remains at an approximate constant energy. The higher
spin members 6$^+_1$, 8$^+_1$
are shifted upwards in a rather regular way down to mass A=198 (N=114), before first the 6$^+_1$ 
starts dropping systematically down in excitation energy at A=196 (N=112), followed by the 8$^+_1$ dropping too 
from A=194 (N=110) (see Figs.~\ref{fig-system-hg} and \ref{fig-energ-comp}) . 
Therefore, it looks like the region 208 $\geq$ A $\geq$ 198 is reminiscent of a vibrational pattern where
the specific two-nucleon properties of the high-spin proton pair remain rather well intact down to the lower
value at A=198. 
An interesting test to gain deeper insight in this region, and the connection to the yet lower mass region
196 $\geq$ A $\geq$ 190 where a dominant rotational-like and more collective pattern is showing up, can be
derived from a study of, e.g., the B(E2;8$^+_1 \rightarrow$ 6$^+_1$)
value, as well
as, of the g-factor and the quadrupole moment of the high-spin 8$^+_1$
state (see Table \ref{tab-be2a} and
Fig.~\ref{fig-state8-1}). Experiments \cite{Stone05,Stone14} have shown
that the g-factor stays remarkably constant going down to A=198, at a value consistent with the
$\pi(1h_{9/2})^2$8$^+$ configuration (for a pure v=2 configuration, as
a function of n, the number of holes in the $\pi(1h_{9/2})$ orbital, the g-factor remains constant), 
indicating a minor influence of collective
admixtures into this state. Maj {\it et al.}~\cite{maj90} have analyzed the B(E2;8$^+_1 \rightarrow$ 6$^+_1$) value as well
as the 8$^+_1$ quadrupole moment in order to extract an effective charge (see e.g., Fig.~9 in \cite{maj90}). The
influence of particle-core coupling on the quadrupole moment of the 8$^+_1$ state has been studied for the Po
nuclei with A=202 and 204 indicating an increase in the quadrupole moment as low as A =200 \cite{ney97}. The
effective charge shows a steady increase of $\sim$ 2 from A=210 down to mass A=200, from which a sharp drop is
observed moving towards A=198.  This seems to support the idea that, in particular for the high-spin members with
a seniority v=2 character, those states only couple weakly with collective quadrupole excitations. Moreover, 
these results also corroborate the fact that from A=196 and onwards with decreasing A value,
large collective components should appear in the 6$^+_1$ state, even
though the g-factor of the 8$^+_1$ remains 
constant between A=200 and A=198 \cite{Stone05,Stone14}. In view of
the discrepancies between the IBM-CM calculations and the experimental
data concerning the g-factor and the B(E2;8$^+_1 \rightarrow$ 6$^+_1$)
value, those states should have a very weak coupling
with the quadrupole collective excitations, and genuinly correspond to
almost pure v=2 seniority configurations (see Fig.~\ref{fig-state8-1}).     

\begin{figure}[hbt]
  \centering
  \includegraphics[width=.6\linewidth]{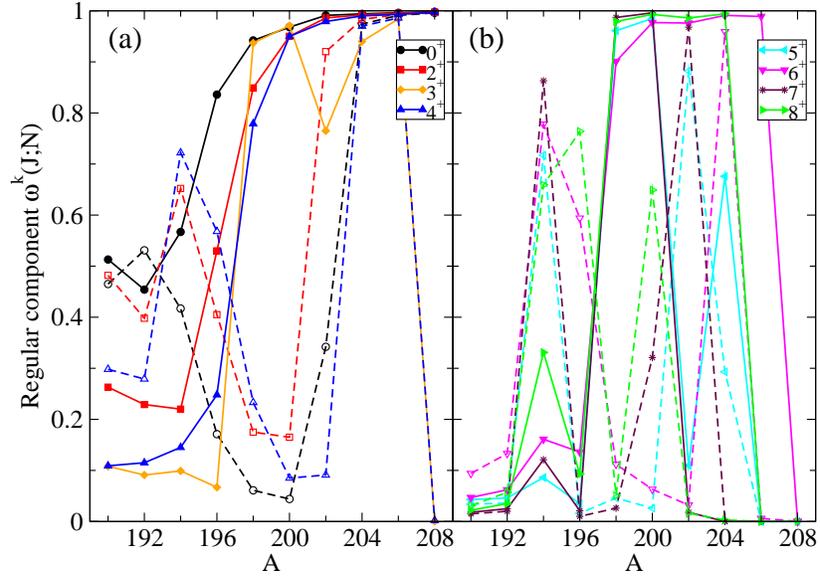}
  \caption{(Color online) Regular content of the two lowest-lying states
    for each $J$ value (full lines with
    closed symbols correspond with the first state while dashed lines with open
    symbols correspond with the second state) resulting from the IBM-CM
    calculation, as presented in figure \ref{fig-energ-comp}.}
  \label{fig-wf}
\end{figure}

Finally, it is worth to note a recent beyond-mean-field calculation
performed for the lead region by Yao {\it et al.}~\cite{yao13} and, in
particular, with results for the even-even Po isotopes
(A=186-204). They have calculated energy spectra  
(general systematics), as well as charge radii,
$\rho^2(0^+_2\rightarrow 0^+_1)$, quadrupole moments for the lowest
$2^+$ and $4^+$ states, and $B(E2;2^+_k \rightarrow 0^+_l)$ (with $k,l=1,2$).
Comparisons with existing data have also been carried out for
$^{190}$Po and $^{194}$Po in Ref.~\cite{grahn08} and for $^{196}$Po in
Ref.~\cite{grahn09}). Because their formulation is within a mean-field
context, invoking the concept of nuclear shapes (spherical, oblate,
prolate), it is not always possible to compare with the present IBM-CM
approach in which the model space is restricted to 0p-0h and 2p-2h
excitations across the Z=82 closed shell. A few general remarks are in place. 
Regarding the
energy spectra the main conclusion of the calculations are the
overestimation of the $0_2^+$ energy (which is predicted to be
oblate), though the slope of the
variation with respect to A is well reproduced and, moreover, that the
calculation is not able to reproduce the almost constant value of the
$2_1^+$ excitation energy, giving rise to a steady increase of the
energy as a function of A. Another very interesting outcome of the
calculation is the spectroscopic quadrupole moment (note that in
\cite{yao13} an equivalent magnitude is provided, $\beta_s$) for the
$2_1^+$ and $2_2^+$ states. The calculation predicts a prolate shape
for the $2_1^+$ state while oblate for the $2_2^+$ state up to N=106 (A=190),
interchanging at this point their character, becoming almost
spherical (but slightly oblate) for N=114 (A=198) and onwards.
In the case of A=190, the calculation exhibits the presence of a
prolate and an oblate band structure, very much like the IBM-CM model
results. The results for the charge radii give an overall correct
trend. However, the upsloping trend in $\langle r^2 \rangle$
(relative to the value for $^{210}$Po), starting at N=112 (A=196) is
not well described within the BMF description as compared with the IBM-CM. 

\subsection{Wave function structure: from the unperturbed structure
  to configuration mixing}
\label{sec-evolution}

We start our analysis with the structure of the configuration-mixed wave functions 
along the yrast levels, expressed 
as a function of the $[N]$ and $[N+2]$ basis states, as given in eq.~(\ref{eq:wf:U5}). 
In Fig.~\ref{fig-wf}a and Fig.~\ref{fig-wf}b, we present 
the weight of the wave functions contained within the $[N]$-boson subspace, defined as
the sum of the squared amplitudes $w^k(J,N) \equiv \sum_{i}\mid a^{k}_i(J;N)\mid ^2$, for
both the yrast states, $k=1$, and the $k=2$ states (the latter are
indicated with a dashed line) for spins $J^\pi=0^+, 2^+, 3^+, 4^+$ in panel (a) and
$J=5^+, 6^+, 7^+, 8^+$ in panel (b).  
The results exhibit an interesting behavior: the yrast states show rather
an intruder character for mass A=190 that quickly are changing into
a regular one in the mass interval A=194-200. For the $0^+$
states, they are fully mixed for A=190 and subsequently, the $0_1^+$  state is changing 
into a pure regular one, while the character of the $0_2^+$ state is essentially
purely intruder for A=200, later on changing into a pure regular
character for A=204. The behavior for the $2^+$ and $4^+$ states is very
similar. The first ($2^+$ or $4^+$) state starts as a dominant 
intruder state for the lighter isotopes, steadily changing with an increase 
of its regular component,  becoming a fully regular configuration for A=200
and onwards to the heavier masses. 
The second $2^+$ and $4^+$ state start with a regular component
of $\approx 50\%$ and  $\approx 30\%$, respectively, reaching a
maximal value for A=194 ($\approx 70\%$), then decreasing when moving  
to A=200, increasing up to reach a purely regular character by mass
A=202. Indeed, the $0_2^+$, $2_2^+$, and $4_2^+$ states exhibit a
similar trend.    
The other states exhibit a more erratic behavior, mainly because of the
many crossings in the unperturbed energy spectra of the regular and intruder states, 
although appearing at very similar energies.  Finally, note that the results for
A=208 should all correspond to pure regular states because the intruder
states appear at a much higher energy at N=126. Because of the constraint 
put on the value of $\Delta^{N+2}$ (see the discussion in section \ref{sec-corr_energy}),
the intruder states appear too low in the energy spectra. 
Moreover, because of the reduced number of bosons in $^{208}$Po (Z=84,
N=124), $2$, only regular states up to spin
J=4 can be constructed within the IBM. Consequently, it is not
 possible to describe the higher-spin (J=8 and beyond) states 
showing up in the mass region 204 $\leq$ A $\leq$ 208 in a reasonable way within the IBM, in view of the specific high-spin
broken-pair states appearing at $\approx$ 1.5-2 MeV.

\begin{figure}[hbt]
  \centering
  \includegraphics[width=.5\linewidth]{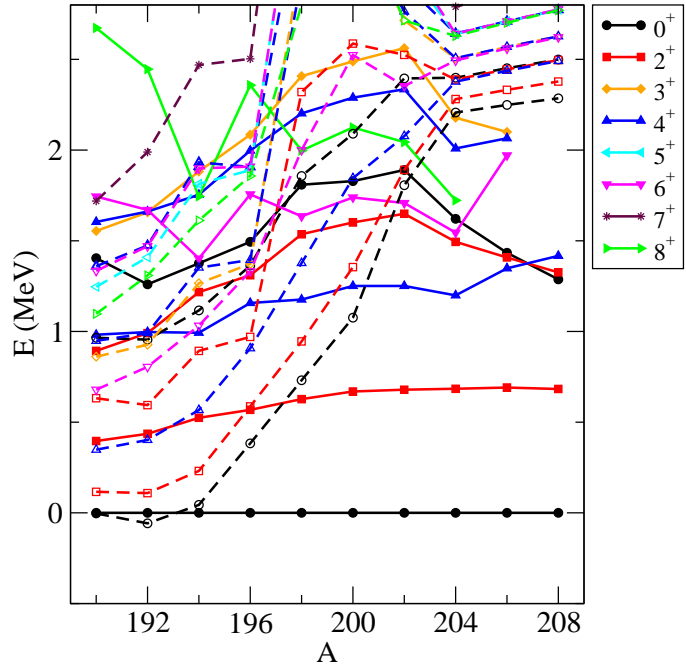}
  \caption{(Color online) Energy spectra for the IBM-CM Hamiltonian presented in Table 
    \ref{tab-fit-par-mix}, switching off the mixing term. The two
    lowest-lying regular states and the
    lowest-lying intruder state for each of the angular momenta are shown (full lines with
    closed symbols for the regular states while dashed lines with open
    symbols are used for the intruder ones).}
  \label{fig-ener-nomix}
\end{figure}

Regarding the energy systematics of the intruder states, one 
expects a parabolic shape centered around N=104, as is the case of Hg
and Pb. However, this is not the case for Po, neither for Pt isotopes. The
reason for not observing the parabolic shape is the rather strong interaction
between regular and intruder configurations and the crossing of the
intruder and regular states in the ground state. Therefore, it is 
very enlightening to calculate the energy spectra as a function of A
switching off the mixing part of the Hamiltonian, i.e. keeping
$w_0^{N,N+2}=w_2^{N,N+2}=0$. These spectra are depicted in Fig.~\ref{fig-ener-nomix} 
where we show the lowest two regular and the lowest two intruder
states for different angular momenta. One observes a rather flat behavior of
the energy for the regular states, still indicating a down-sloping
tendency when moving towards the lighter Po isotopes. The energy of the intruder
states is smoothly decreasing down to neutron mid-shell (N=106). 
In this region, the parabola become very flat (A=190-194). 
This results mainly from the smooth change of the Hamiltonian
parameters when passing from isotope to isotope. A striking fact
is the almost degeneracy of the unperturbed regular and intruder $0^+$ states for
A=190-194, with the intruder configuration becoming the lowest one in the energy spectrum for
A=190-192. The crossing of a regular and a intruder configuration with the
same angular momentum has a strong influence on the regular component
of the states resulting from the full IBM-CM calculation (see
Fig.~\ref{fig-wf}) inducing an interchange of character between the two
states. In the situation of the $J^{\pi}=2^+$ unperturbed energy spectrum, the closest approach happens at A=196
and in Fig.~\ref{fig-wf} one notices that the $2_{1,2}^+$ states
interchange their character at this point. The same happens for the $4^+$ states at A=196-198.

A most interesting decomposition of the wavefunction is obtained by first calculating the wavefunctions within the N subspace as
\begin{equation}
\Psi(l,JM)^{reg}_N = \sum_{i} c^{l}_i(J;N) \psi((sd)^{N}_{i};JM)~, 
\label{eq:wf:N}
\end{equation}
and likewise for the intruder (or N+2 subspace) as
\begin{equation}
\Psi(m,JM)^{int}_{N+2} = \sum_{j} c^{m}_j(J;N+2)\psi((sd)^{N+2}_{j};JM)~,
\label{eq:wf:N+2}
\end{equation}
defining an ``intermediate'' basis~\cite{helle05,helle08}. 
This generates a set of bands within the 0p-0h and 2p-2h subspaces, 
corresponding to
the unperturbed bands that are extracted in schematic two-level phenomenological model calculations 
(as discussed in references ~\cite{bree14,liam14,kasia15,duppen90,drac88,drac94,allatt98,page03,drac04}),
and indeed correspond to the unperturbed energy levels depicted in Fig.~\ref{fig-ener-nomix}.

\begin{figure}
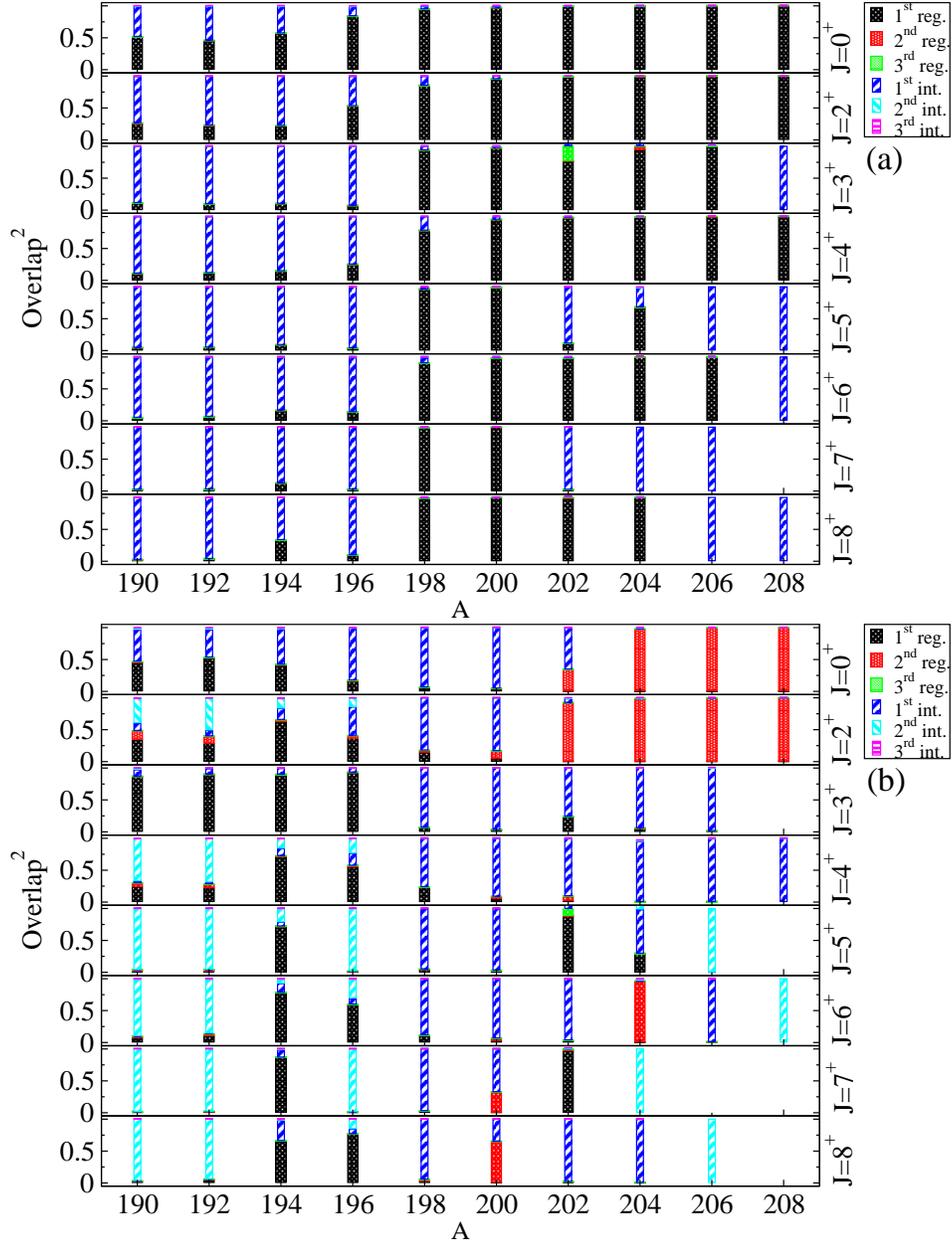

\includegraphics[width=0.7\textwidth]{overlap-int-reg-1.eps}\\
\includegraphics[width=0.7\textwidth]{overlap-int-reg-2.eps}%
\caption{(Color online) Overlap of the wave functions of Eq.~(\ref{eq:wf:U5}), with
  the wave functions describing the unperturbed basis  
Eq.~(\ref{eq:wf:N}) and Eq.~(\ref{eq:wf:N+2}). Panel (a): overlaps
for first $0^+,2^+,3^+,4^+;5^+,6^+,7^+,8^+$ state,  
  panel (b): overlaps for the corresponding second state (see also text).}
\label{fig-overlap} 
\end{figure}

The overlaps $_{N}\langle l,JM \mid k,JM\rangle$ and $_{N+2}\langle m,JM \mid k,JM\rangle$
can then be expressed as,
\begin{equation}
_{N}\langle l,JM \mid k,JM\rangle=\sum_{i} a^{k}_i(J;N) c^{l}_i(J;N), 
\end{equation} 
and  
\begin{equation}
_{N+2}\langle m,JM \mid k,JM\rangle=\sum_{j} b^{k}_j(J;N+2) c^{m}_j(J;N+2),
\end{equation} 
(see expressions (\ref{eq:wf:N}) and (\ref{eq:wf:N+2})).
In Fig.~\ref{fig-overlap} we show these overlaps, but squared, where we restrict
ourselves to the first and second state ($k=1,2$) with 
angular momentum J$^{\pi}$=0$^+$,2$^+$,3$^+$,4$^+$,5$^+$,6$^+$,7$^+$,8$^+$,
and give the overlaps with the lowest three bands within the
regular $(N)$ and intruder $(N+2)$ spaces ($l=1,2,3$ and $m=1,2,3$).  Since
these figures are given as a function of mass number, one obtains a
graphical insight into the changing wave function content. 
In panel (a), which corresponds to the first state, one observes for
the $0^+$ state a strong mixing between the first regular
and first intruder unmixed states (also called the ``intermediate basis'') for A=190-194, while for the
heaviest isotopes the state is mainly of regular character. For the
$2^+$ states the situation is very much the same though the mixing is
much more reduced. For the
higher spin states one observes a sharp border at A=198, in such a
way that for A $\leq 196$ the states correspond to the first intruder
configuration, while for A $\geq 198$ they correspond to the first regular
configuration. Note that the intruder character observed for the heavier
isotopes is somehow artificial because of the reduced number of bosons
and also to the reduced excitation energy of the intruder configurations.
In panel (b) the situation becomes rather fuzzy especially for $J>2$. For the
$0^+$ state one observes a mixture between the first regular and intruder
configurations up to A=196, then the composition changes into a pure first intruder configuration and
finally ends up as the second regular configuration. Concerning the $2^+$ state, the
lighter isotopes correspond to a mixture of the second intruder and
the first and second regular configuration, the composition changing into the first
intruder for A=198-200, finally ending as the second regular configuration. For the higher-spin
states, the pattern is not so clear. In general, the
states become rather pure and the changes in their composition is mainly due to the
many crossings between the unperturbed regular and intruder configurations.     

\section{Study of other observables: alpha-decay hindrance factors and isotopic shifts}
\label{sec-other}

\subsection{$\alpha$-decay hindrance factors}
\label{sec-alpha}

The $\alpha$-decay process can be used as an important tool to probe nuclear structure, in particular
the overlap factor between the initial and final wave functions has the character of an $\alpha$-
particle spectroscopic factor. Because calculations of the absolute decay rate are very difficult, most
often one studies decay branches, characterized by a hindrance factor described by the expression  
\begin{equation}
HF = \frac{\delta_{gs}^2}{\delta_{ex}^2}=\frac{I_{gs}P_{ex}}{I_{ex}P_{gs}},
\end{equation} 
where $\delta_{i}^2$ is the reduced $\alpha$ width, $P_{\alpha_i}$ the
penetration probability through the combined Coulomb 
and centrifugal barrier \cite{duppen00} and $I_i$ the $\alpha$-decay intensity
(with $i= gs, ex$ for the ground state and excited state, respectively) \cite{duppen00}.

The Pb-region has been studied intensively and hindrance factors have been used to obtain valuable 
information on shape coexistence, and mixing of various nuclear shapes. In particular, a two-level
mixing model, suggested by Wauters {\it et al.}~\cite{wauters94a}, has been used to analyze the experimental 
hindrance factors and extract complementary information on mixing between different nuclear configurations.
\begin{figure}[hbt]
  \centering
  \includegraphics[width=.4\linewidth]{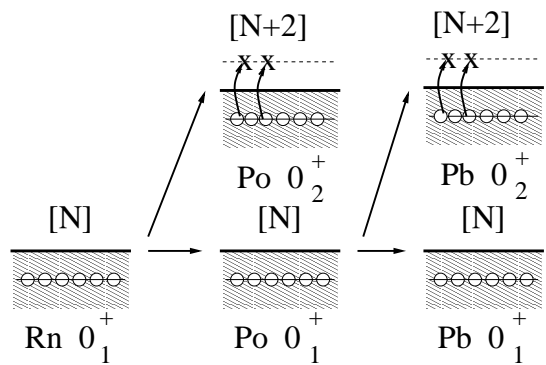}
  \caption{Schematic view of $\alpha$ decay from the Rn nuclei into the Po nuclei, 
   and from the Po nuclei into the Pb nuclei.}
  \label{fig-sch-alpha}
\end{figure}

The structure information on the composition of the $0^+_1$ ground state in the $^{190-198}$Po nuclei is
derived from the specific $\alpha$ decay hindrance factors into both the $0^+_1$ and $0^+_2$ states into the
corresponding daughter $^{186-194}$Pb nuclei (see Fig.~\ref{fig-sch-alpha}). Use is made of the fact that for the Pb nuclei, it was
shown, through study of the Pb nuclear charge radii \cite{heyde11} and very weak E0 decay strength 
from the $0^+_2 \rightarrow 0^+_1$ transtion, that the Pb nuclei retain a spherical ground-state structure.
The experimental results on these Po nuclei \cite{waut93,waut94,wauters94a,bijn96,allatt98,andrey99a,duppen00,huyse01} have been analyzed
in these papers resulting in hindrance factors of $2.8\pm 0.5$,
$2.5\pm 0.1$, $1.1\pm 0.1$, and $0.4$ for A=198, 196, 194, and 192, respectively.  
The major result is the decreasing trend in these hindrance factors, even
favoring decay to the excited $0^+_2$ excited state at $591$ keV in $^{188}$Pb starting from $^{192}$Po. 
In those studies, and using a two-state mixing model \cite{wauters94a}, an intruder character for 
the ground state $0^+_1$ state in $^{192}$Po turns out to be $\approx$ 63\% \cite{allatt98}, or, even
larger $>65\%$ \cite{andrey99a}. More recently, $\alpha$-decay from $^{188,192}$Po has
been observed into the final $0^+$ excited states in $^{184,188}$Pb \cite{vande03}. Their analyses results in hindrance
factors into the first excited $0^+_2$ state of $0.08 \pm 0.03$ and $0.57 \pm 0.12$, respectively, which
may well turn out as in indication of a very large overlap of the wavefunctions in initial and excited
state in the final nucleus. In that latter paper, an analyses was carried out from inspecting the various
energy minima in both the Po and Pb nuclei, obtaining qualitative information on overlaps of
the corresponding wave functions, derived from the Nilsson-Strutinsky approach \cite{satula95}. An analysis
of hindrance factors was also carried out using a similar method, making use of deformed mean-field wave
functions at the energy minima of the total energy surfaces, pointing out that this approach, albeit different
from an approach explicitely including mp-nh excitations across the closed Z=82 shell closure, renders quite
similar resuls \cite{karl06}.

The wavefunctions, obtained from the IBM-CM (see also similar results on the study of the Pt \cite{Garc11} 
and Hg \cite{Garc14} nuclei on $\alpha$-decay hindrance factors), 
give rise to the following theoretical values of
$6\%, 17\%, 42\%, 53\%$, corresponding with the weight factors of the intruder component (see Fig.~\ref{fig-wf}) for
the masses A=198, 196, 194, and 192, respectively. These numbers can be
compared with the experimental relative hindrance factors as discussed before, with a rather good agreement.

On the other side, $\alpha$-decay starting from the even-even $^A$Rn
nuclei (see Fig.~\ref{fig-sch-alpha}), into the $^{A-4}$Po nuclei, gives very
strong fingerprints for observing excited $0^+$ states, in particular, experiments studying the decay of $^{202}$Rn into $^{198}$Po
\cite{waut92} and of $^{198,200}$Rn\cite{bijn95} into the $^{194,196}$Po nuclei. The analysis of these data,
using a two-state mixing model ~\cite{duppen90}, resulted in detailed information on the
more deformed intruder configuration in the $2^+_1$ state. Its weight factor  
is changing from $7\%$, passing by $31\%$ and
$72\%$, towards the value of $88\%$, for A=200, 198, 196, 194, respectively. It is interesting to compare these results
with the present IBM-CM calculations resulting in a variation of the intruder weight factor of $5\%$, $15\%$, $47\%$, and 
$78\%$ (see Fig.~\ref{fig-wf}) for the same mass values. This again indicates a rather good description of the changing wave function
structure for the $2^+_1$ state in the Po even-even nuclei. 

Studies on $\alpha$-decay from the odd-mass Po nuclei have been carried out for the same mass region, with
now an extra neutron coupled to the underlying Po even-even core \cite{andrey99,andrey02,andrey06,vande05}.
These experiments show the presence of rather weakly particle-core coupled states, associated with a spherical
configuration for the even-even underlying core, as well the appearance of a more strongly coupled structure,
associated with coupling of the odd particle with the more deformed configuration of the even-even core.
\begin{figure}[hbt]
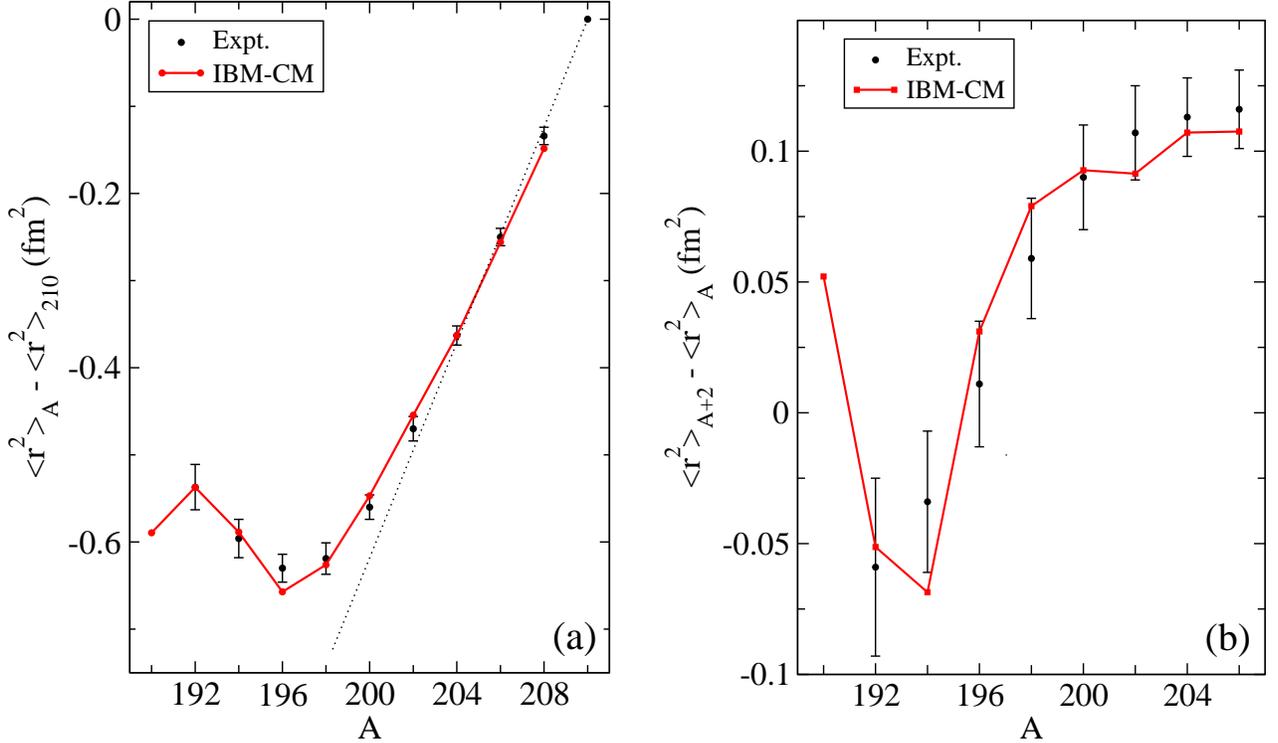

  \centering
  \includegraphics[width=.45\linewidth]{radius.eps}
~~~~
  \includegraphics[width=.45\linewidth]{iso-shift.eps}%
  \caption{(Color online) Panel (a): Charge
    mean-square radii for the Po nuclei. Panel (b): Isotopic shift for the Po
    nuclei.  The data are taken from \cite{cocio11}.}
  \label{iso-shift}
\end{figure}

A lot of effort has been invested into going beyond the more simplified two-state mixing model that has been 
extensively studied
analysing the $\alpha$-decay hindrance factors related to the decay into the excited $0^+_i$ states in the
daughter nuclei. In particular, 
more detailed calculations have been carried out
by Delion {\it et al.}~\cite{delion95,delion96}, and more recently in \cite{delion14},
emphasizing the need for a microscopic QRPA description that encompasses both nedutron and 
proton pairing vibrations as well as including proton 2p-2h excitations across the $Z=82$ closed shell.
More in particular, calculations have mainly concentrated in order to better understand the observed hindrance factors 
for $\alpha$-decay leading into the neutron-deficient Po, Pb, Hg, and Pt nuclei.  
Karlgren {\it et al.}~\cite{karl06} use the information on the energy minima 
in the $\beta-\gamma$ plane of the total energy surface as the essential input to calculate the hindrance
factors. They also discuss a possible connection between the present deformed mean-field approach with a picture
in which specific particle-hole excitations are invoked within a shell-model approach.  
Moreover, Xu {\it et al.}~\cite{xu07} formulates the $\alpha$-decay hindrance factors within a density-dependent
cluster model to describe the tunneling of a preformed cluster through the deformed Coulomb barrier.

\subsection{Isotopic shifts}
\label{sec-radii}

Experimental information about ground-state charge radii is also available
for both the even-even \cite{cocio11} and odd-mass \cite{seliv13,seliv14} Po nuclei.
Combined with similar data for the adjacent
Po, Pb, and Pt nuclei, as well as for the odd-mass Bi, Tl and Au nuclei, the systematic
variation of the charge radii supplies invaluable
information on the ground-state wave function \cite{ulm86,otten89,kluge03}. 
We illustrate the overall behavior of ${\langle r^2 \rangle}_A$ relative
to the radius at mass A=210 in Fig.~\ref{iso-shift}a and 
the relative changes defined as
$\Delta {\langle r^2 \rangle}_A \equiv \langle r^2 \rangle_{A+2}$ -$\langle
r^2 \rangle_{A}$ in Fig.~\ref{iso-shift}b. The experimental data are 
taken from Cocolios {\it et al.}~\cite{cocio11}. 

To calculate the isotope shifts, we have used the standard IBM-CM expression for the nuclear radius
\begin{equation}
r^2=r_c^2+ \hat{P}^{\dag}_{N}(\gamma_N \hat{N}+ \beta_N
\hat{n}_d)\hat{P}_{N} + 
\hat{P}^{\dag}_{N+2}(\gamma_{N+2} \hat{N}+ \beta_{N+2} \hat{n}_d) \hat{P}_{N+2}.
\label{ibm-r2}
\end{equation} 
The four parameters appearing in this expression are fitted 
to the experimental data, corresponding to the radii of A=192-198 
  even-even Po isotopes \cite{cocio11}. The resulting values
are $\gamma_N=-0.108$ fm$^2$, $\beta_N=0.152$ fm$^2$,
$\gamma_{N+2}=-0.022$ fm$^2$, and $\beta_{N+2}=0.027$ fm$^2$. Note
that in determining these parameters we took as reference point $^{204}$Po
instead  of $^{210}$Po which is the reference for the experimental data.

Panel (a) of  Fig.~\ref{iso-shift} shows the value of the radius
relative to A=210. The first important fact is the strong deviation
from the linear trend (marked with the dotted line) at A=198 and
downwards which is showing the onset of deformation confirmed
in several recent experimental papers \cite{seliv13,cocio11}. To see
this fact more clearly, we also present the value of the isotopic
shift which enhances the appearance of the deformation. We mention 
the very good quantitative agreement between the IBM-CM and the
experimental data, which confirms that the interplay between
the $[N]$ and $[N+2]$ contributions in the $0^+$ ground state wave
function along the whole chain of Po isotopes is well described by the
model.
Comparing with the data for the nearby Hg and Pt isotopes, one notices
that the range of variation of the isotopic shift for Hg is only
$\approx 0.02$ fm$^2$, while for Pt this amounts to $\approx 0.1$ fm$^2$
and, finally, for Po becomes $\approx 0.2$ fm$^2$. This fact is telling us  
how abrupt the onset of deformation in the Po is. Note the strong
similarity of Fig.~\ref{iso-shift}b with Pt case \cite{Garc11}, where there
is also an abrupt drop of the isotopic shift at A=184-186. 
\begin{figure}
\includegraphics[width=0.6\textwidth]{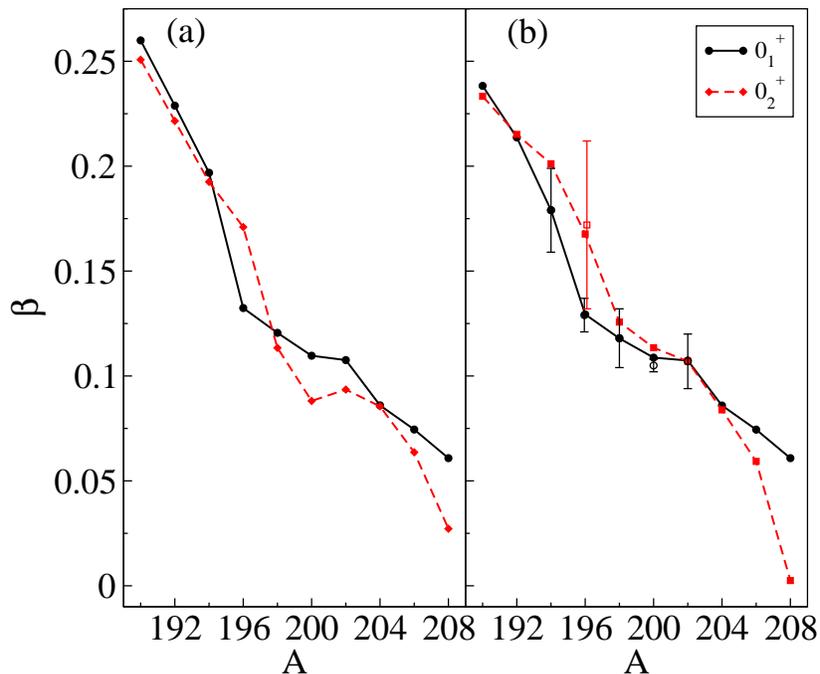}
\caption{(Color online) Comparison of the value of $\beta$ extracted from the
  theoretical quadrupole shape invariants (panel (a)) and the ones
  extracted from experimental and theoretical B(E2) values (panel (b)),
  corresponding to the $0_1^+$ and $0_2^+$ states.} 
\label{fig_beta}
\end{figure} 

\section{Quadrupole invariants and nuclear deformation}
\label{sec-q-invariants}

\begin{figure}
\includegraphics[width=0.3\textwidth]{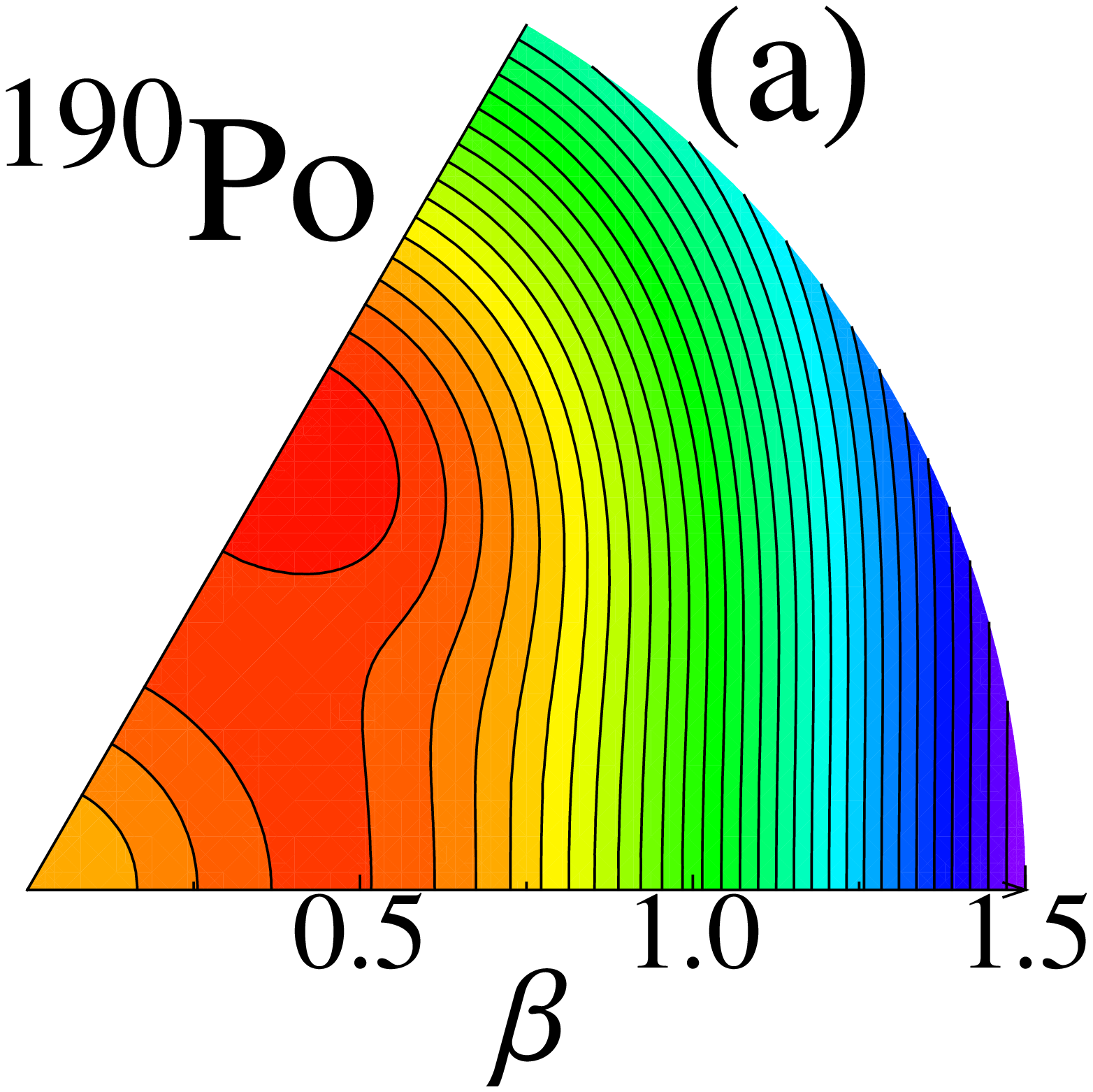}%
\includegraphics[width=0.3\textwidth]{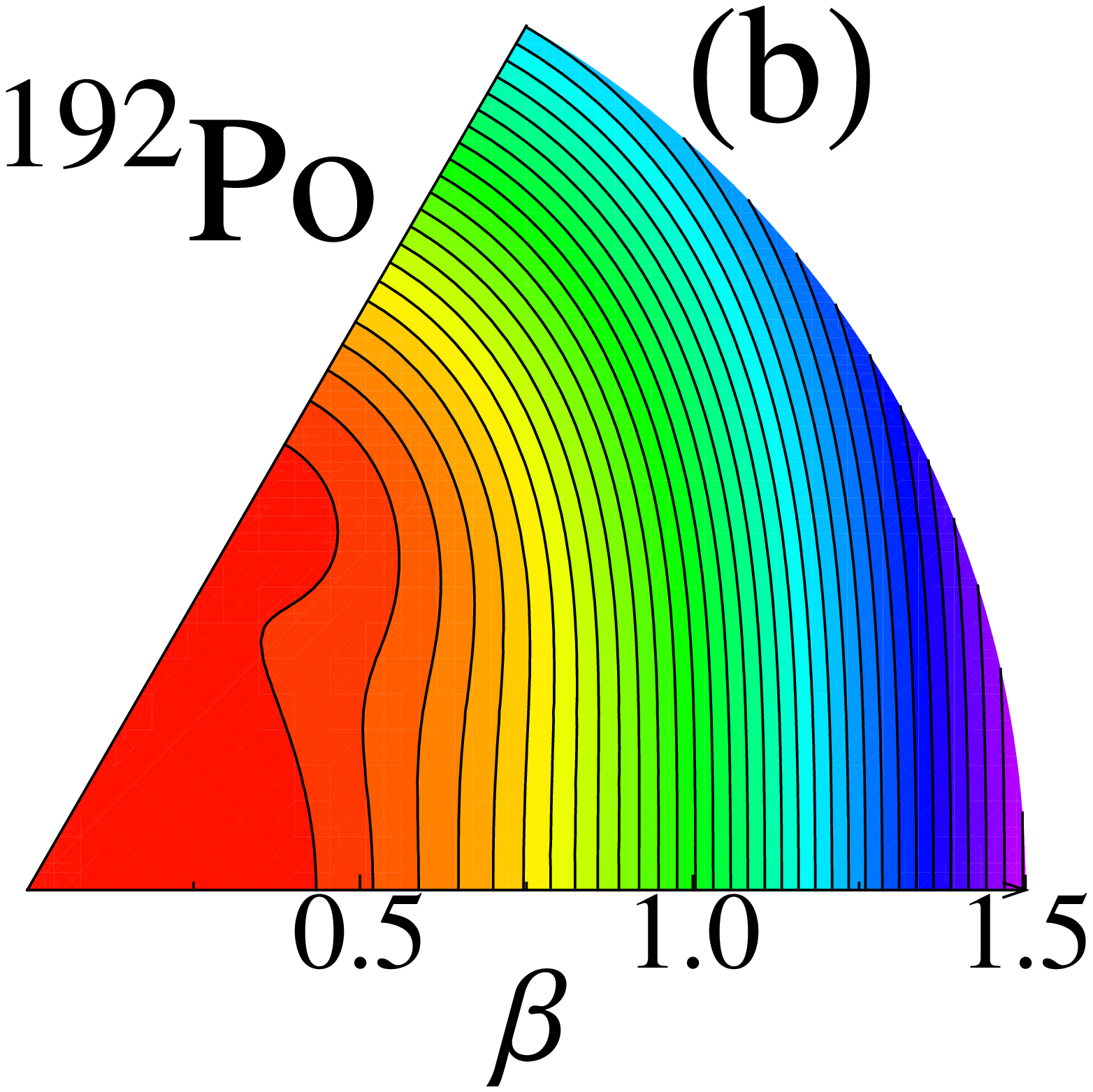}%
\includegraphics[width=0.3\textwidth]{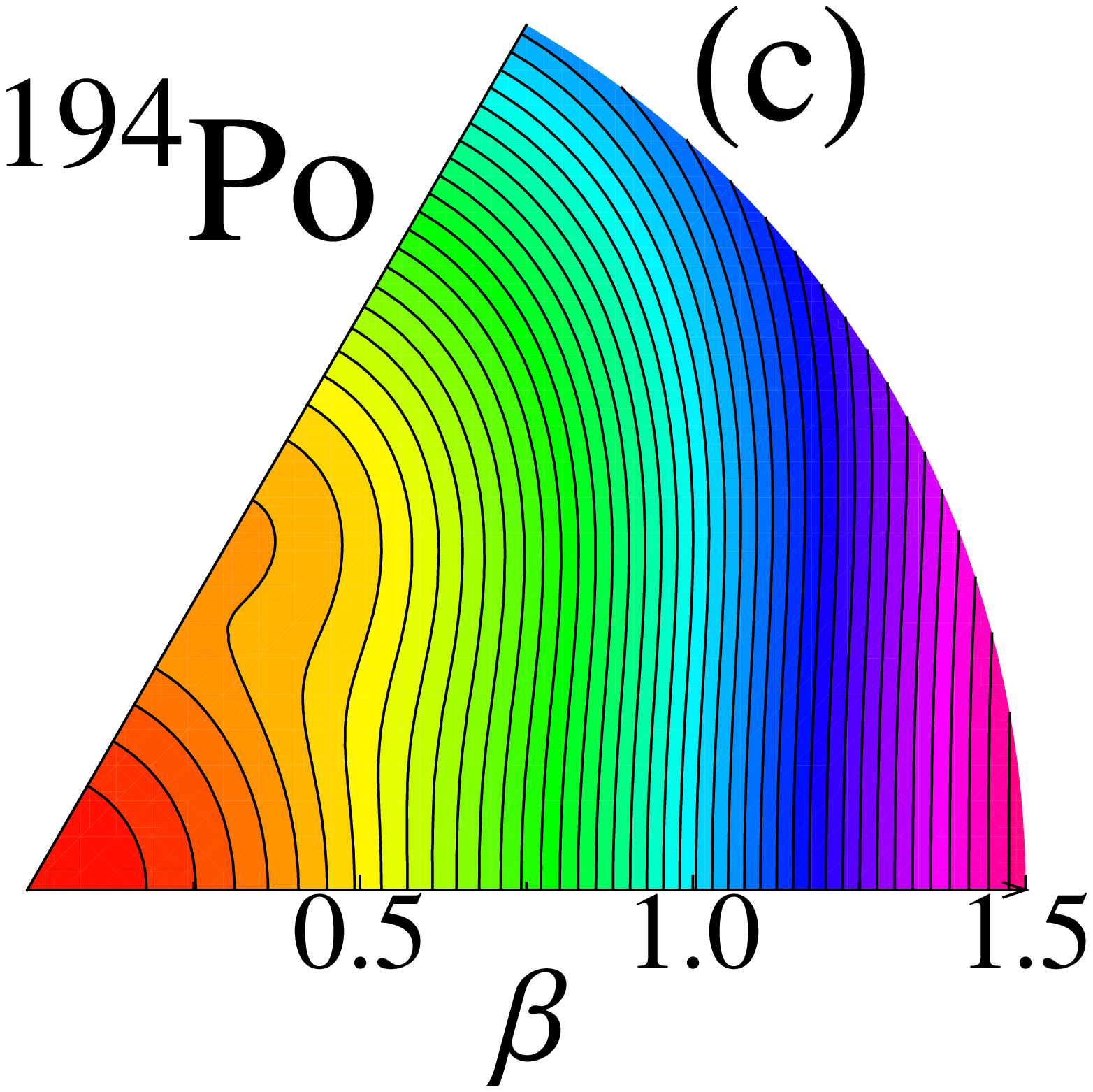}
\caption{(Color online) Matrix coherent-state calculation for $^{190-194}$Po,
  corresponding with the present IBM-CM Hamiltonian (table
  \ref{tab-fit-par-mix}). The energy spacing  between adjacent contour lines equals $100$ keV and the
  deepest energy minimum is set to zero, corresponding to the red color.}
\label{fig_ibm_ener_surph}
\end{figure}

The IBM can provide us with both the energy spectra and the corresponding wave functions, as well
as all derived observables (B(E2)'s, quadrupole moments, radii, ...),
working within the laboratory frame, as well as the 
corresponding mean-field energy surface, defining a nuclear shape over
the $\beta-\gamma$ intrinsic frame.  This geometric interpretation of
the IBM can be constructed using the 
intrinsic state formalism, as proposed in
Refs.~\cite{Bohr80,gino80,diep80a,diep80b,gil74}. The extension of this
formalism to describe
simultaneously regular and intruder configurations was proposed by
Frank {\it et al.,} introducing a matrix coherent-state method 
\cite{Frank02,Frank04,Frank06,Mora08}. In Refs.~\cite{Garc14a,Garc14b}
a detailed description of the method and its application to Pt and Hg
isotopes, respectively, can be found.  
In Figs.~\ref{fig_ibm_ener_surph}a, \ref{fig_ibm_ener_surph}b, and
\ref{fig_ibm_ener_surph}c the IBM-CM mean-field energy surfaces are
depicted for $^{190}$Po, $^{192}$Po, and $^{194}$Po, respectively. In
$^{190}$Po the energy surface presents an oblate minimum, that evolves
into a very flat surface for $^{192}$Po and in a spherical shape for
$^{194}$Po. Note that from Fig.~3 in Ref.~\cite{yao13}, the $^{192}$Po
and $^{194}$Po nuclei  
exhibit an oblate minimum, however, in reaching $^{190}$Po, the energy
minimum switches over to become a prolate minimum, with a very similar
deformation. 

Even though the shape of the nucleus is not an experimental observable, it is still possible to extract
from the data direct information about various moments characterizing the nuclear shape corresponding
with a given eigenstate. Using Coulomb excitation, it is possible to extract the most important diagonal
and non-diagonal quadrupole and octupole matrix elements, including their relative signs and, in a model
independent way, extract information about nuclear deformation  
as shown by Kumar, Cline and co-workers \cite{kumar72,Cline86,Wu96,clement07,sreb11}. 

The essential point is the introduction of an ``equivalent ellipsoid'' view of a given nucleus ~\cite{kumar72}
corresponding to a uniformly charged ellipsoid with the same charge, $\langle r^2 \rangle$, and 
E2 moments as the nucleus characterized by a specific eigenstate \cite{kumar72,wood12}. 

From the theoretical point of view the nuclear shape can be calculated using the quadrupole shape
invariants. The quadrupole deformation corresponds to
\begin{eqnarray} 
q_{2,i}&=&\sqrt{5} \langle 0_i^+| [\hat{Q} \times \hat{Q} ]^{(0)}|0_i^+
\rangle.
\label{q_invariant1}
\end{eqnarray} 
For the triaxial rigid rotor model~\cite{wood09} it is directly connected with the
deformation parameter
\begin{eqnarray} 
q_2 &=& q^2\label{q_invariant2b},
\label{q_invariant2}
\end{eqnarray} 
where $q$ denotes the nuclear intrinsic quadrupole moment. 

\begin{table}
\begin{ruledtabular}
\caption{Quadrupole shape invariants for the $^{190-208}$Po
  isotopes.} 
\label{tab-q-invariant}
\begin{tabular}{cccccc}
Isotope & State &$ q^2 $ (e$^2$~b$^2$) & Isotope & State &$ q^2 $ (e$^2$~b$^2$)  \\
\hline
$190$ & $0_1^+$ &6.2\footnotemark[1] & $200$ & $0_1^+$ &1.1\footnotemark[1] \\
      & $0_2^+$ &5.7\footnotemark[1] &       & $0_2^+$ &0.9\footnotemark[1] \\
$192$ & $0_1^+$ &4.8\footnotemark[1] & $202$ & $0_1^+$ &1.1 \footnotemark[1]\\
      & $0_2^+$ &4.5\footnotemark[1] &       & $0_2^+$ &0.9\footnotemark[1] \\
$194$ & $0_1^+$ &3.6 & $204$ & $0_1^+$ &0.7\footnotemark[1] \\
      & $0_2^+$ &3.5 &       & $0_2^+$ &0.7\footnotemark[1] \\
$196$ & $0_1^+$ &1.7 & $206$ & $0_1^+$ &0.6 \footnotemark[1]\\
      & $0_2^+$ &2.8 &       & $0_2^+$ &0.4\footnotemark[1] \\
$198$ & $0_1^+$ &1.4 & $208$ & $0_1^+$ &0.4\footnotemark[1] \\
      & $0_2^+$ &1.2 &       & $0_2^+$ &0.1\footnotemark[1] \\
\end{tabular}
\end{ruledtabular}
 \footnotetext[1]{The effective charges have been taken from
   neighboring isotopes (see Table \ref{tab-fit-par-mix}).}
\end{table}

To calculate analytically the quadrupole shape invariants characterizing the nucleus in its
ground-state and low-lying excited states, it is necessary to resort to a closure relation,
$\textbf{1}=\sum_{J,i,M}| J_i M\rangle \langle J_i M|$, leading to the expression
\begin{eqnarray} 
\label{q_invariant4}
q_{2,i} &=&\sum_r \langle 0_i^+|| \hat{Q}|| 2_r^+\rangle  \langle
2_r^+||\hat{Q} ||0_i^+ \rangle.
\label{q_invariant4b}
\end{eqnarray} 

In Table \ref{tab-q-invariant}, we present the theoretical value of
$q^2$ corresponding to the $0_1^+$ and $0_2^+$ states, for the whole chain of Po
isotopes.  In this case, only very few experimental values of the reduced E2 matrix elements 
$\langle 0_i^+|| \hat{Q}|| 2_r^+\rangle$ have been measured \cite{keste15} besides the strongest
and dominating $\langle 0_1^+|| \hat{Q}|| 2_1^+\rangle$ matrix element. 
Because of the limited number of reduced E2 matrix elements, going into r=2,3,.. final states,
it is not possible to extract reliable experimental values. For the
same reason it is not possible to extract  the triaxial shape
variable $\delta$, this is the reason for not including this variable
in the discussion.
The main conclusion of this table is that
the deformation of both 0$^+_{1,2}$ states is very similar and is dropping when moving
from mid-shell, at N=104, to the end of the shell at N=126 \cite{keste15}.    

Starting from the quadrupole invariant (\ref{q_invariant1}), one can extract a value of the deformation $\beta$
(see, e.g., references \cite{Srebrny06,clement07,Wrzo12})

\begin{equation}
\beta=\frac{4\, \pi\, \sqrt{q^2}}{3\, Z\, e\, r_0^2\, A^{2/3}},
\label{beta-q2}
\end{equation}
where $e$ is the proton charge and $r_0=1.2$ fm. Thus, we can extract values
for $\beta$ corresponding to the ground-state $0_1^+$ and the first excited $0_2^+$ state.

The resulting $\beta$ values, extracted from (\ref{beta-q2}) are shown in Fig.~\ref{fig_beta}(a), and one notices an overall
decrease, albeit with a plateau in the mass region 194 $\leq$ A $\leq$ 202, which is the
region corresponding with an overall change in the energy scale for the first excited $2^+_1$ state.

We also provide a different measure of the deformation value and
extract a value of $\beta$ starting from a given B(E2) value 
through the expression:
\begin{equation}
\beta=\displaystyle{\frac{4 \, \pi \, \sqrt{B(E2;J\rightarrow
      J-2)}}{3\, Z\, e \, r_0^2\,
  A^{2/3}\, \langle J\, 0\, 2\, 0\,| J-2\, 0\,\rangle}},
\label{beta-e2}
\end{equation}
where $\langle j_1 m_1 j_2 m_2| j m \rangle$ is a Clebsch-Gordan coefficient.

\begin{figure}
\includegraphics[width=0.6\textwidth]{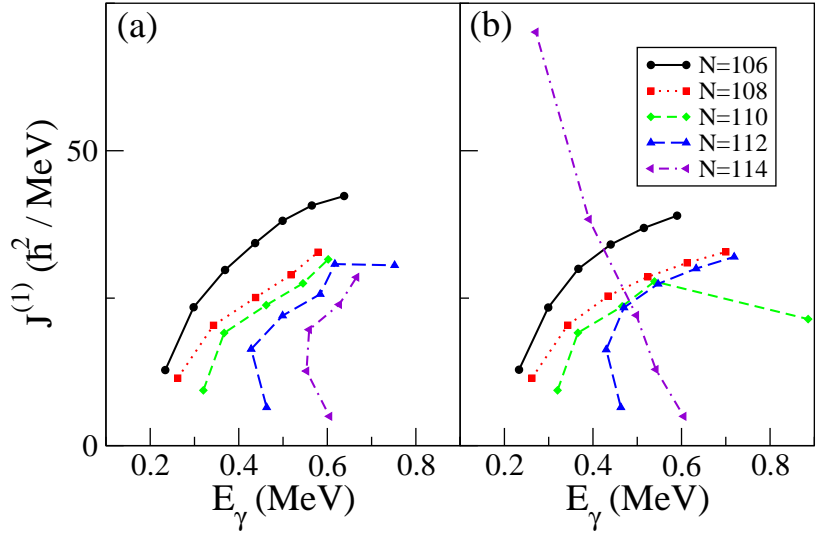}
\caption{(Color online) Comparison of the experimental (a) and
  theoretical (b) kinematic moment of inertia for selected Po isotopes with 106 $\leq$ N $\leq$ 114.} 
\label{fig_mi-po}
\end{figure} 

In particular, we extract the value of $\beta$, derived from the $B(E2;2_1^+\rightarrow 0_1^+)$ value, but also for any other
$B(E2;J^+\rightarrow (J-2)^+)$ value along the yrast band. The extracted $\beta$ values then give
interesting information about a possible variation along the
band. Note that we extract the value for 
$0_1^+$ from $B(E2;2_1^+\rightarrow 0_1^+)$ and for $0_2^+$ from $B(E2;6_1^+\rightarrow 4_1^+)$.   
The corresponding values are plotted in
Fig.~\ref{fig_beta}(b) and are fully consistent with the results,   
as shown in Fig.~\ref{fig_beta}(a). They are slightly smaller because of using only a single but dominant
$B(E2;2_1^+\rightarrow 0_1^+)$ reduced transition probability.

\begin{figure}
\includegraphics[width=0.8\textwidth]{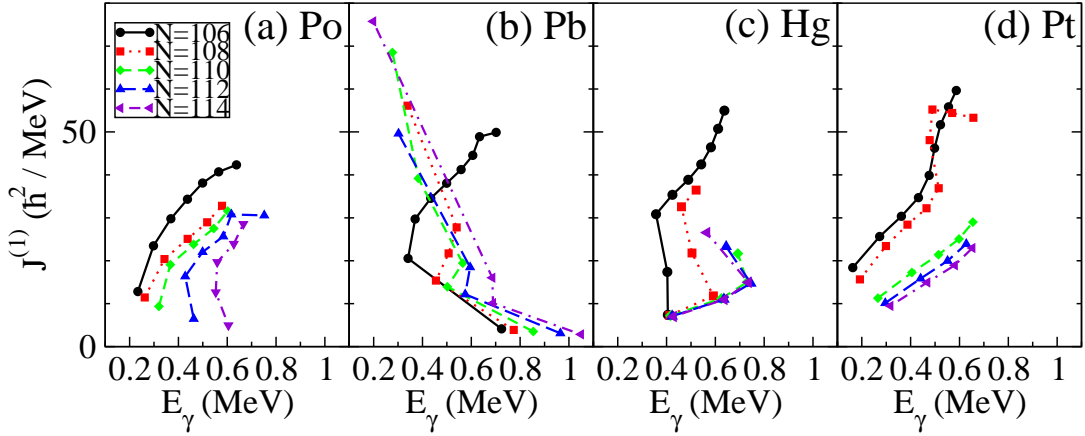}
\caption{(Color online) Comparison of the experimental moment of
  inertia in Po (a), Pb (b), Hg (c) and Pt (d), for selected isotones with 106 $\leq$ N $\leq$ 114.} 
\label{fig_mi-exp}
\end{figure} 

The results for the Po nuclei are contrasting with the nearby Hg and Pt isotopes. In a similar analysis
as the one carried out here, but for the Hg isotopes (see Figs.~2 and 3 in reference \cite{Garc15}), the $\beta$ values extracted from
the IBM-CM results exhibit a strong difference in the interval 180 $\leq$ A $\leq$ 190. This is consistent with the
$\beta$ values extracted from the $B(E2;2_1^+\rightarrow 0_1^+)$ (showing the less deformed oblate structure) and from the
$B(E2;6_1^+\rightarrow 4_1^+)$ (associated with a more deformed prolate structure) and quite different from the overall 
smooth drop in the $\beta$ values extracted in the Po isotopes. Comparing with the Pt nuclei, in which fewer data were
available (see figure 3.27 in \cite{wood92}), down to mass A=188, $\beta$ values extracted from the 
$B(E2;2_1^+\rightarrow 0_1^+)$ and the $B(E2;6_1^+\rightarrow 4_1^+)$ reduced E2 transition probabilities do not
show the same large separation as in the case of the Hg isotopes. In the Pt case, a rather
smooth transition (a bump) from the region with A $\geq$ 188 moving
towards the mid-shell region exists. This was shown 
to be a consequence of the rather strong mixing between regular and intruder configurations \cite{Garc09,Garc11}. 
The behavior of $\beta$, within the context of the IBM-CM, indicated a similar bump as a function of 
mass number \cite{Garc14a}.

Interesting information about the changing collective structure along the yrast band for the Po can be extracted
studying the variation of the moment of inertia as one moves up the band structure. A quantitative measure, that is
often used, is the kinematic moment of inertia \textit{J}$^{(1)}$, which is defined as \cite{bomot75,nils95}
\begin{equation}
J^{(1)}
=\frac{\hbar^2}{2}\left(\frac{dE}{d\left(J(J+1)\right)}\right)^{-1}
\approx \frac{\hbar^2(2J-1)}{2 E_\gamma(J \rightarrow (J-2))} , 
\label{kinetic}
\end{equation}
where $E_\gamma(J \rightarrow (J-2))$ is the energy diffence $E(J)-E(J-2)$.
For an ideal rigid rotor, with a moment of inertia that is independent of the angular momentum J, the kinematic moment
of inertia reduces to a constant.

In constructing the experimental \textit{J}$^{(1)}$ values, we use the yrast band energies for the corresponding J$^{\pi}$ values.
However, approaching the high-spin levels with 8$^+$, 10$^+$,12$^+$,
it can happen that non-collective (mainly broken-pair configurations)
do cross the collective structure, giving also rise to isomeric states
as a fingerprint. Consequently, 
considering those states, the smooth variation of \textit{J}$^{(1)}$ will change into a backbending pattern. Experimental data on
lifetimes, $\gamma$-ray intensities characterizing the decay within a band structure, as well the coincidence data 
with $\gamma$-rays from the low-energy transitions in the experimental
yrast bands, have been used to construct the resulting \textit{J}$^{(1)}$ values. 
In particular, for the Po nuclei (see references
\cite{jul01,wise07,hela96,grahn09,bern95,lach94}), we cover the  106 $\leq$ N $\leq$ 114 (190 $\leq$ A $\leq$ 198), region, 
with the results presented in Fig.~\ref{fig_mi-po}. A similar analysis spanning a smaller
set of Po nuclei, was carried out in references \cite{jul01,hela99,vande03a,wise07}. 
One observes a very smooth
increase along the yrast band, moving up to rather high spin values
in the A=190 (up to spin 14$^+$), 192 and 194 (up to spin 10$^+$) isotopes, consistent with
the energy spectra as shown in Fig.~\ref{fig-system-hg}. It is only for the higher-spin states, which are of a non-collective broken-pair
nature, that very small energy differences appear, not at all consistent with collective excitations and giving rise to
strong back-bending (not shown here). 
From A=198, and onwards, a different energy pattern  for \textit{J}$^{(1)}$ is observed.
For the lower part of the band, up to the 6$^+_1$ level, the values of E$_{\gamma}$  
are roughly constant, indicating the typical energy differences of a vibrational structure.
The fairly correct reproduction of this trend by the IBM-CM results indicates that the major physics content is rather 
well described.

It is also interesting to compare - making use of the experimental data on band structure and shape coexistence in the Pb region -
the Po nuclei, with the neighboring Pb, Hg and Pt isotopes in order to appreciate similarities and differences, covering the
same neutron interval 106 $\leq$ N $\leq$ 114, as shown in
Fig.~\ref{fig_mi-exp}. The data used to construct the
\textit{J}$^{(1)}$ band for the Po, Pb, Hg, and Pt isotopes have 
been taken from the references given in Section \ref{sec-exp} and \cite{drac98}.  

An obvious observation from Fig.~\ref{fig_mi-exp}, is the fact that at N=106, i.e. 
the nuclei $^{190}$Po,$^{188}$Pb,$^{186}$Hg and $^{184}$Pt, from 
J$^{\pi}$= 6$^+$ onwards, a very similar variation in the value of \textit{J}$^{(1)}$ results. The starting point in these nuclei is 
$\approx$ 30 $\hbar^2$ MeV$^{-1}$, followed by a similar slope up to spins 12$^+$ and 14$^+$. In view of the well-established
prolate character in the case of $^{188}$Pb and $^{186}$Hg, it is a convincing argument that the ground-state band in both $^{190}$Po
and $^{184}$Pt, from spin 4$^+$, 6$^+$, and onwards, behave like a prolate band too. 
One also notices that since the ground state 0$^+$ and first excited 2$^+$ state in $^{188}$Pb retain mainly a spherical character,
the low E$_{\gamma}$ part exhibits a different sloping start, very much like it is the case in $^{186}$Hg. In the latter nucleus, there appears
an off-set of the prolate  band versus the ground-state less deformed band (oblate, anharmonic vibrational) causing
almost constant values of \textit{J}$^{(1)}$ to result for the 2$^+$,4$^+$ states. In $^{190}$Po and $^{184}$Pt, a very smooth behavior results 
along the yrast band, extending up to rather high-spin states and this for the whole region 106 $\leq$ N $\leq$ 114. 
One also observes a considerable drop
in the value of \textit{J}$^{(1)}$ in this interval, which is of the order of $\approx$ 10 $\hbar^2$ MeV$^{-1}$. 
In this respect, the Po and Pt nuclei exhibit 
a rather similar collective behavior, with the "intruder" states being not that obvious from the experimental data on
excitation energies nor the corresponding B(E2) electromagnetic properties. Only radii and alpha-decay hindrance
factors point towards the need to consider both configuration (regular [N] and intruder [N+2]) spaces.  
 
\section{Conclusions and outlook}

\label{sec-conclu}

The Po region constitutes one of the clearer examples where shape
coexistence plays a key role in explainig the major observed features such as the
evolving energy spectra, with mass number decreasing from the neutron closed
shell at N=126, approaching the mid-shell region at N=104, the electromagnetic properties,
in particular the B(E2) systematics, nuclear mean-square charge radii, $\alpha$-decay
hindrance factors, etc. These isotopes are situated in the region of the Pb, Hg and Pt isotopes  
where the proximity of the Z=82 proton closed shell, and its stabilizing effect to keep nuclei
mainly into a spherical shape, in particular for the Pb nuclei, is very
evident.
These series of isotopes can be divided in two groups: on the one
hand, the Pb and Hg nuclei, in which two (or even three for the Pb nuclei) types of
configurations coexist and are experimentally well documented, characterized by energy spectra that exhibit
the characteristic parabolic behavior of the intruder band structure, minimizing its excitation energy at 
the mid-shell, N=104 \cite{fossion03,Garc14a}. 
On the other hand, the Pt and Po isotopes for which one cannot
disentangle easily the presence of two distinct band structures. Still, in both the Pt and Po
isotopes, approaching the mid-shell point, the first excited 0$^+$ state behaves in an unexpected way, dropping
seriously in excitation energy which can be used as an indirect hint for the presence of shape coexistence, 
however in a concealed way \cite{Garc11}.

In the present paper, we have carried out an extensive study of the chain of isotopes
$^{190-208}$Po using the Interacting Boson Model, including pair boson excitations across the Z=82 proton closed
shell (called the intruder configuration space), and their interaction with the regular configuration space, 
called the IBM-CM. We have determined the Hamiltonian and the E2 operator describing this interacting system 
of bosons, through a least-squares fit to the known experimental data. This then results in the energy spectra, and,
moreover, the calculation of many different observables such as
the B(E2) values, nuclear mean-square charge radii, gyromagnetic factors and 
$\alpha$-decay hindrance factors give the possibility to test the nuclear dynamics. 
In particular, the latter three observables are shown to serve as
fingerprints to test the relative composition of the nuclear wave function into its "regular" and "intruder"
components, and thus are strong indicators for the presence of shape coexisting structures.

A very important issue 
concerns deriving information on the nuclear deformation properties. In particular, recent Coulomb excitation
experiments on nuclei far from stability made it possible to extract a set of reduced E2 matrix elements
$\langle 0^+_i\mid\mid\hat{T}(E2)\mid\mid 2^+_f\rangle$. These matrix elements can be used to 
construct the so-called "quadrupole shape invariants" for the 0$^+_1$ and 0$^+_2$ states. 
We have given particular attention to how $\beta$-values can be derived from these invariants, or alternatively,
from known experimental B(E2) values, and compared both approaches in the case of the Po nuclei.
Thereby, a clear picture of the shape coexistence phenomenon in
Po arises. There are two families of configurations, one slightly deformed (rather
$\gamma$-unstable) or spherical that corresponds to the regular
configurations, while the other more deformed and corresponding to the
intruder configurations. Because of the lack of experimental information it is
not possible to determine unambiguously the shape connected to the two families of
configurations and, indeed, through the use of mean-field results (see
section \ref{sec-theo}) 
we assume an oblate shape for the intruder
states, although the moment of inertia seems to suggest a prolate
shape for the ground state of $^{190}$Po. Both configurations have a rather large interplay and in the
mid-shell region with the two $0^+$ unperturbed bandheads being degenerate in energy implying important
mixing between the two configuration spaces.
We have found that, in contrast with the Hg nuclei, the difference in the quadrupole deformation extracted for
the 0$^+_1$ and 0$^+_2$ states is quite small and the mass dependence very similar, independent of the method used
to extract the quadrupole deformation.

Concluding, it looks like the data available at present, based on our present study of the Po isotopes and a 
comparison with the Pt isotopes points towards rather similar structures. 
In both cases the ground-state at the mid-shell is composed mainly from the intruder configurations, 
although in the Po isotopes, only the second part of the neutron shell
N=82-126 has been studied experimentally. For the Pt isotopes, almost the whole shell
has been covered, but for the Po nuclei, because of the sparse character of the known data
in and below N=106, A=190, a detailed comparison is very difficult. 
Moreover, the similar character of the Po and Pt isotopes is supported from a comparison of the kinematic moments of
inertia for the Po and Pt isotopes: both exhibit a smooth variation as a function of neutron number in the
interval 106 $\leq$ N $\leq$ 114. Comparing with the corresponding moments of inertia in the Pb and Hg isotopes,
the nuclei $^{190}$Po, $^{188}$Pb, $^{186}$Hg, and $^{184}$Pt (starting at spin 6$^+$) are very similar, resulting in the suggestion
that from that spin, and onwards, the $^{190}$Po ground band behaves like a prolate band. This is further corroborated by
a comparison of the known experimental excitation energies for the prolate bands that have been 
observed in $^{186}$Hg and in $^{188}$Pb
for states with spin from 6$^+$ to 10$^+$. They match very well the corresponding energies in the yrast band structure
in $^{190}$Po.
Possibilities to test this proposal might come partly from multiple Coulomb excitation with higher energy projectile ions
(HIE-ISOLDE), and, possibly at a later stage, using one and two nucleon transfer reactions and studying E0 properties for the Po nuclei,
which may lead to the direct observation of a low-lying excited 0$^+$ state, as suggested by the present study.

\section{Acknowledgment}
We are very grateful to
N.~Kesteloot for generous sharing of their most recent results on
Coulomb excitation (A=196-198).
We thank M. Huyse,
P.~Van Duppen for continuous interest in this research topic and
J.L.~Wood for stimulating discussions in various stages of this work.
Financial
support from the ``FWO-Vlaanderen'' (KH and JEGR) and the InterUniversity
Attraction Poles Programme - Belgian State - Federal Office for
Scientific, Technical and Cultural Affairs (IAP Grant No.  P7/12) is
acknowledged.  This work has also been partially supported by the
Spanish Ministerio de Econom\'{\i}a y Competitividad and the European
regional development fund (FEDER) under Project No.
FIS2011-28738-C02-02 and by Spanish Consolider-Ingenio 2010
(CPANCSD2007-00042).

\end{document}